Title: **Deciphering autism heterogeneity: a molecular stratification approach in four mouse models**

Caroline GORA[1], Ana DUDAS[1], Océane VAUGRENTE[1], Lucile DROBECQ[1], Emmanuel PECNARD[1], Gaëlle LEFORT[1] and Lucie P. PELLISSIER[1]*

[1]INRAE, CNRS, Université de Tours, PRC, 37380, Nouzilly, France

*corresponding author: Lucie P. Pellissier, PhD, Team biology of GPCR Signaling systems (BIOS), INRAE, CNRS, Université de Tours, PRC, 37380, Nouzilly, France. Phone: +33 4 47 42 79 62. Email: lucie.pellissier@inrae.fr

Running title: **Defining markers for the stratification of autism**

**ABSTRACT**

Autism spectrum disorder (ASD) is a complex neurodevelopmental condition characterized by impairments in social interaction, communication, as well as restrained or stereotyped behaviors. The inherent heterogeneity within the autism spectrum poses challenges for developing effective pharmacological treatments targeting core features. Successful clinical trials require the identification of robust markers to enable patient stratification. In this study, we explored molecular markers within the oxytocin and immediate early gene families across five interconnected brain structures of the social circuit in four distinct ASD mouse models, each exhibiting unique behavioral features along the autism spectrum. While dysregulations in the oxytocin family were model-specific, immediate early genes displayed widespread alterations, reflecting global changes in social plasticity. Through integrative analysis, we identified *Egr1*, *Foxp1, Homer1a*, *Oxt* and *Oxtr* as five robust and discriminant molecular markers facilitating successful stratification of the four models. Importantly, our stratification demonstrated predictive values when challenged with a fifth mouse model or identifying subgroups of mice potentially responsive to oxytocin treatment. Beyond providing insights into oxytocin and immediate early gene mRNA dynamics, this proof-of-concept study represents a significant step toward potential stratification of individuals with ASD. The implications extend to enhancing the success of clinical trials and guiding personalized medicine for distinct subgroups of individuals with autism.

## INTRODUCTION

Autism spectrum disorder (ASD) is a heterogeneous neurodevelopmental condition, with a prevalence of 1% [1]. ASD is characterized by impairments in social communication and interaction, as well as repetitive or stereotyped behaviors [2], associated with various co-occurring features. Despite the identification of over thousands of genes associated with ASD, the precise etiology of this complex condition remains predominantly idiopathic. The help provided to individuals with ASD, addressing their core social features, heavily relies on intensive and costly early behavioral interventions [3, 4]. While medications have gained approval for targeting co-occurring features, their efficacy in improving life-quality of people with ASD is often limited, coupled with potential side effects [5, 6]. Notably, the administration of atypical antipsychotics alleviates anxiety, self-injury, and compulsive behavior among individuals with ASD [5, 6]. To date, clinical trials for pharmacological interventions in autism have encountered setbacks, with a lack of available drugs specifically targeting core features [7, 8]. The failure of these trials can be attributed to the considerable heterogeneity observed across the spectrum, particularly in larger cohorts involved in phase 2b/3 trials. This inherent diversity within the patient population underscores the critical necessity for stratification within the ASD spectrum. The success of future trials hinges on our ability to effectively delineate subgroups of individuals with autism, necessitating the identification of reliable markers.

The neuropeptide oxytocin (OT), along with its paralog vasopressin (AVP), and their corresponding receptors play a pivotal role in regulating social behaviors across the lifespan [9–12]. These neuropeptides are predominantly synthesized in neurons of the paraventricular (PVN) and supraoptic (SON) nuclei of the hypothalamus [13–16] and project extensively to brain structures involved in emotional and social circuit, such as the prefrontal cortex and striatum, where their receptors are expressed, or the pituitary [17–19]. Furthermore, there is emerging evidence suggesting a potential link between the oxytocin system and the etiology of ASD (for reviews [20–23]). Pathogenic variants identified in oxytocin (*OXT*), oxytocin receptor (*OXTR*), and vasopressin receptor 1A (*AVPR1A*) genes have been associated with ASD (SFARI Gene). Notably, the deletion of these genes in mice has been shown to induce autism-like traits (for review [8]). Nevertheless, the administration of OT has yielded inconsistent effects in individuals with ASD [8, 24, 25]. Intriguingly, a study has demonstrated that the concentration of OT in the blood can predict the outcome of oxytocin administration [26], suggesting a potential role for OT as a marker for stratification of individuals diagnosed with autism. However, it is crucial to note that OT peptides are rapidly released in both the plasma and the brain within minutes, triggered by the activation of OT neurons in response to various stimuli, including social and non-social cues [27–31], as well as stressful conditions [32–36]. Consequently, the focus should shift from the OT peptide itself to its mRNA, which may better reflect long-term synaptic plasticity impairment. Indeed, post-mortem studies have consistently reported decreased *OXT* and *OXTR* mRNA levels within the brains of individuals with ASD or related psychiatric conditions [37, 38], as well as in mouse models of ASD [23, 39]. The crucial question of whether mRNAs within the OT family can serve as common or specific molecular markers for people with ASD  remains unanswered.

Individuals with ASD exhibit global alterations in synaptic plasticity, as evidenced by multiple studies [40–43]. Synaptic plasticity, a vital adaptive process responding to various stimuli, including social interactions [44–46], plays a pivotal role in shaping neural circuits. Immediate early gene (IEG) mRNAs constitute a key component of this process, as they are rapidly transcribed, transported, and translated within 30 minutes to 2 hours in response to diverse stimuli. Following this initial response, IEGs orchestrate long-term plasticity by initiating the second wave of late genes within 4 to 24 hours. Notably, IEGs emerge as central factors potentially influencing impairments in social interactions. Dysregulation of the expression levels of IEGs has been observed in both plasma and brain tissues of people with ASD [47–51]. These dysregulations are not limited to clinical observations, as animal models of ASD also display altered IEG expression [52–54]. Whether specific IEGs can serve as predictive markers for distinct autism subtypes remains an open avenue for investigation.

In this study, we used three distinct genetic mouse models — *Shank3*, *Fmr1* and *Oprm1* knockout (KO) mice — alongside mice subjected to early chronic social isolation, encompassing diverse etiologies to replicate the autism spectrum. Our investigation revealed that each model exhibited a distinctive behavioral signature, mimicking the complexity of the spectrum. We pinpointed *Oxt*, *Oxtr*, *Foxp1*, *Egr1*, and *Homer1a* as robust molecular markers within the OT and IEG gene families. These markers exhibited specific responses to social interactions in wildtype (WT) mice and significant dysregulations in the aforementioned four mouse models across five structures of the social circuit. Using these five markers, we successfully demonstrate the first proof-of-concept for the stratification of mouse models, differentiating subtypes within the spectrum of autism. The identification of these specific markers offers a promising avenue for tailoring personalized medical help to individuals with autism.

## MATERIALS AND METHODS

### *Animals*

All mouse breeding, care and experimental procedures were in accordance with the European and French Directives and approved by the local ethical committee CEEA Val de Loire N°19 and the French ministry of teaching, research and innovation (APAFIS #18035-2018121213436249). Two months old socially naive males and females from *Oprm1* KO (JAX stock #007559) [55], *Shank3* KO (JAX stock #017688) [56], *Fmr1* KO and females *Fmr1*$^{+/-}$ (provided by Rob Willensem; [57]), *Arc* KO (JAX stock #007662) [58] and WT animals were maintained on a 12-hour light/dark regular cycle, with food and water *ad libitum* and controlled temperature (21°C) and humidity (50%). Mice were raised in the mixed C57BL/6J;129S2 background in social groups of 3-4 individuals. Animals named “isolated”, were exposed to chronic social isolation after weaning and for at least 4 weeks prior to behavioral tests.

### *Behavior experiments*

All behavioral tests were sequentially performed in a dim light quiet room, using standardized equipment and are detailed in the **supplementary methods**. Briefly, social interactions were performed in the 3-chambered and in the reciprocal social interaction tests with a sex- and age-matched unknown mouse or a cage mate in an open field. Non-social interactions were tested with an object also in open field arenas. Perseveration and cognitive flexibility in a spatial task was assessed in the Y-maze test while repetitive and stereotyped behaviors were examined in the motor stereotypy test, assays that have been previously characterized for mouse models of ASD [59].

### *Quantitative PCR*

Mice were dissected at basal conditions or 0.75 (45 minutes), 2 or 6 hours following a social or an object interaction. Within 5 minutes, 1 mm thick brain slices were generated using a coronal mouse brain matrix. Prefrontal cortex (PFC), nucleus accumbens (NAC), caudate putamen (CPU), paraventricular (PVN) and supraoptic (SON) nuclei (**Figure S1E**) were collected using a 2 mm diameter puncher (two punches for lateralized regions) and immediately frozen until further use. Total RNAs were extracted according to the ZymoResearch's instructions (Direct-zol™ RNA Microprep and Miniprep kit). The cDNAs were generated from 0.5 µg or 0.25 µg of total RNAs using the SensiFast reverse transcriptase kit and quantitative PCR (qPCR) was performed in triplicates according to ONEGreen Fast® kit with 1 µL of cDNA and 1 µM of each validated couple of primers (**Table S1**). The following qPCR protocol was applied for 40 cycles: 95°C for 5s, 60°C for 15s, and 60°C for 30s. All materials, reagents, protocols and suppliers are detailed in **Table S2**.

### *Data modeling and correlations*

All statistical analyses were performed using R (version 4.2.2) [60]. Animal outliers (+/- 3 standard deviations) were removed. Variability between cohorts (**Figure S1**) was assessed using Principal Component Analysis (PCA) with the FactoMineR package [61], and for qPCR, corrections were applied using the ComBat sva package [62]. For behavioral data, Kruskal-Wallis tests with Dunn's *post hoc* tests were conducted using the rstatix package [63]. P-values were adjusted with Benjamini-Hochberg correction [64]. A complete linear model was fitted for qPCR data, encompassing variables such as time, social interaction or mouse line, and their interactions. Post-hoc tests, based on estimated marginal means with Tukey p-value correction, were conducted using the emmeans package [65] to compare social interaction at each time point, the different time point for each interaction, and mouse models to WT mice. In addition, the analyses were conducted separately as PCA revealed distinct clusters among the structures: one comprising PFC, NAC, and CPU, and the other involving PVN and SON (**Figure S1**). Additionally, in WT mice, clustering of genes and median expressions was performed using the pheatmap package. Furthermore, integration of qPCR and behavioral data was performed using DIABLO (Multiblock (s)PLS-DA) implemented in the mixOmics package [66, 67]. The compromise parameter was set at 0.75 to maximize correlation between qPCR and behavioral datasets. Linear discriminant analysis (LDA) was performed using *Egr1*, *Foxp1*, *Homer1a, Oxt* and *Oxtr,* on each mouse with less than 5 missing values across all structures, using the MASS package [68]. LDA predicted the class membership of *Arc* KO mice that were not considered during model fitting (*i.e.*, the model with most similarities). Finally, on the data of *Shank3* and *Fmr1* KO mice and only on the OT markers, hierarchical clustering on the principal components of PCA (HCPC) was performed using the FactoMineR package [61].

**RESULTS**

**Distinct behavioral profiles across autism spectrum in four mouse models**

We examined the behavioral profiles of *Fmr1* and *Shank3* KO mice, modeling Fragile X and Phelan-McDermid syndromes, respectively, along with *Oprm1* KO and chronically isolated mice (**Figures 1**, **S2-S3**, **Table S3**), all previously published as mouse models of ASD and neurodevelopmental conditions [56, 69–71]. To capture a comprehensive view of their social responses, we exposed these models to distinct social interactions: one involving a family member (e.g. a cage mate, "SI mate") and the other with an unknown conspecific ("SI unknown"), encountered in the daily life of individuals with autism. Additionally, we included a non-social interaction condition with an object ("NSI object"), as control.

In the open-field, none of the models exhibited altered behavior following SI mate or NSI object (**Table S3**). *Fmr1* KO mice demonstrated robust social impairments, indicated by a decrease in both total time and mean duration engaged in nose contacts with an unknown mouse across two distinct tests (**Figure 1A, D**). This impairment was also observed in heterozygous females in the sociability phase of the three-chambered test (**Figure S2A-C**). In contrast, *Shank3* KO mice displayed impaired social novelty, evident in their lack of preference for a new mouse over a familiar one (**Figure 1E**). Surprisingly, *Oprm1* KO and isolated mice did not exhibit the expected social impairments as reported previously [69, 71]. Under standard conditions, *Oprm1* KO mice only displayed a lack of mate preference in the three-chambered test (**Figure 1F**). Elevating light intensity from 15 to 40 lux to slightly increase anxious-like behavior revealed social interaction impairments with an unknown animal in *Oprm1* KO mice, as evidenced by a reduction in the time spent in nose contact during both the sociability and social novelty tests, as opposed to WT mice tested in dim light conditions (**Figure S2D-F**). This finding suggests a potential manifestation of induced social impairments or social anxiety — a phenotype situated at the periphery of the autism spectrum.

Regarding stereotyped behaviors, *Shank3* KO mice spent more time in self-grooming, accompanied by an increased number of head shakes and a decrease in the time spent digging and in the number of rearing events compared to WT mice (**Figures 1B-C**, **S3A-B**). Conversely, *Oprm1* and *Fmr1* KO mice demonstrated a reduction in self-grooming time, while isolated mice exhibited a decreased number of head shakes (**Figure 1B-C**). Concerning co-occurring features, none of the models exhibited impaired cognitive flexibility in the Y maze, nor did they show locomotion impairments evidenced by no differences in the traveled distance (**Figure S3C-D**). Notably, *Shank3* KO mice displayed anxious-like behaviors, spending more time in the periphery of the open field arena compared to WT mice (**Figure S3E**).

In summary, the four models presented distinct behavioral features, with *Fmr1* and *Shank3* KO mice displaying the most severe phenotypes along the autism spectrum.

**Identification of specific molecular markers of social interactions in WT mice**

Given that molecular markers for social interactions remain unknown, our primary objective was to identify them, starting with WT mice. We focused on two mRNA families, recognized as potential markers of autism [8, 23, 42, 72]: the OT family, encompassing *Oxt* and *Avp*, and their receptors (*Oxtr*, *Avpr1a*, *Avpr1b*), and IEGs and neurotrophic factors, including *Arc*, *Egr1*, *Fos, Fosb, Jun*, *Homer1a*, *Foxp1, Gdnf* and *Bdnf*. We unraveled their kinetic profiles up to 6 hours after SI mate, SI unknown, and NSI object, as well as 6 hours of acute social isolation across five key structures within the social circuit: the PVN, SON, CPU, NAC, and PFC. To ensure the specificity of our findings, we ruled out circadian or sex biases within the two mRNA families. Notably, none of these mRNAs displayed distinct patterns within a 6-hour period or between males and females, except for *Foxp1*, which exhibited sexually dimorphic expression in the SON (**Figure S4A**, **Table S4**).

Markers were selected based on significant differences observed between one SI and NSI object, between SI unknown and SI mate, as well as differences between time points exclusively for one SI (**Figures S5-S7**). Minimum criterion of significance in at least two different structures was applied. Consequently, we identified key markers, specifically *Avp*, *Avpr1a, Oxt* and *Oxtr* mRNAs within the OT family, along with *Arc, Egr1, Fos*, *Fosb, Foxp1* and *Homer1a* within the IEG family. Notably, SI unknown emerged as the most distinct social stimulus, revealing 26 significant differences across the five structures, with 21 differences observed at 45 minutes (**Figures 2**, **S5-S7**, **Table S4**). This stimulus triggered a rapid and transient increase or decrease, followed by normalization to basal levels at 2 and 6 hours. In contrast, SI mate demonstrated 14 distinct and sustained regulations, with 8 differences evident at 2 and 6 hours. No single structure or mRNA distinctly stood out for a particular social stimulus (**Figure 2**).

*Avp* and *Oxt* peptide mRNAs exhibited pronounced differences between the two social stimuli in the PVN and SON, where the somas of OT and AVP neurons are located (**Figures 2**, **S5**). Specifically, SI mate rapidly elevated *Avp* and *Oxt* mRNA levels that lasted over 6 hours. Conversely, SI unknown increased transiently *Avp, Oxt,* and *Oxtr* expression at 45 minutes in the NAC and CPU, as well as *Avpr1a* at 2 hours and 45 minutes, respectively (**Figure 2**). Remarkably, our datasets revealed a robust positive correlation, reaching up to 0.92, between *Oxt* and *Avp* levels in WT mice, with the PVN exhibiting the lowest correlation at 0.85 (**Figure**

S8). *Avpr1b* mRNA levels remained unchanged, irrespective of time and the nature of social interaction. These findings highlight the intricate and context-dependent regulatory mechanisms governing the expression of peptide and receptor mRNAs within the OT family.

Among the IEG family, *Egr1* and *Foxp1* emerged as the most discriminant markers of social interactions, exhibiting significant differences across all five structures, followed by *Arc, Fos* and *Homer1a* in four structures and *Fosb* in three structures (**Figures 2**, **S6-S7**). Notably, *Egr1, Fos*, *Foxp1* and *Homer1a* showed a high induction at 45 minutes following SI unknown in the PVN and SON (**Figure 2**). *Foxp1* stood out as a distinct IEG, revealing reduced levels at 45 minutes following SI unknown in the CPU and PFC. Interestingly, *Arc* in the SON was the only mRNAs down-regulated by acute social isolation from cage mates, a condition with negative social valence (**Figure S4B**). Despite the interesting profile of *Fosb*, which mirrored *Egr1*, as evident from their robust correlation (**Figure S8C**), we included only its relative *Fos*. *Bdnf, Gdnf* and *Jun* mRNAs did not meet the selection criteria, exhibiting responses in only one structure or none. Noteworthy, *Foxp1*, *Homer1a* and *Egr1* displayed a negative correlation with *Oxt* and *Avp* (**Figure S8C**), suggesting a potential interplay between them.

In summary, we identified *Arc*, *Avp, Avpr1a, Egr1, Fos*, *Foxp1*, *Homer1a, Oxt* and *Oxtr* as specific molecular markers of social interactions in WT mice. These markers were selected to assess their dysregulations in the four mouse models, both under basal conditions and following SI unknown at 45 minutes — the most discriminant interaction and time point.

## Distinct dysregulations within the oxytocin family among mouse models

Previous studies have proposed OT, AVP, or their receptors as potential common biomarkers of autism [8, 23, 72]. To explore shared dysregulations within the OT family, using *Oprm1, Fmr1*, *Shank3* KO and isolated mice, we assessed alterations in *Oxt*, *Avp, Oxtr* and *Avpr1a* expression following SI unknown at 45 minutes and under basal conditions (**Figures 3**, **S9-S10**, **Table S5**). In the CPU, *Avp* was not induced across all four mouse models, as well as *Oxtr* in three models, compared to WT animals (**Figure 3A**). Additionally, we identified two shared dysregulations between two models — *Avpr1a* in the SON and *Oxt* in the NAC (**Figures S9-S10**). However, the majority of dysregulations were rather specific for each mouse model. *Fmr1* KO mice displayed a global decrease in the expression of all four mRNAs in the PVN, along with *Avp*, *Oxt*, *Oxtr* in the NAC, and *Oxt* and *Avpr1a* in the SON (**Figures 3B**, **S9-S10**). In contrast, isolated mice exhibited an overall increase in *Avp, Oxt* and *Oxtr* expression in the PFC (**Figure 3C**). *Shank3* KO mice did not show additional dysregulations, while in the CPU, *Oxt* and *Avpr1a* were also down-regulated in *Oprm1* KO mice (**Figure S9**).

To explore the potential for shared dysregulations beyond the initial OT family, we extended our assessment to include *Cd38,* involved in OT secretion, and *Avpr1b*, along with generalist enzymes involved in peptide biosynthesis (*Pcsk1, Pcsk2, Pcsk5* and *Cpe*) and degradation (*Lnpep*, *Ctsa* and *Rnpep*; **Figure S11**, **Table S5**). Strikingly, only three shared dysregulations — *Cpe*, *Ctsa*, *Rnpep* in the NAC — were observed among two models (**Figure S11**). Once again, the majority of dysregulations demonstrated model specificity. In addition to the shared dysregulations, *Fmr1* KO mice exhibited 6 unique down-regulations, particularly in the NAC and PVN. Isolated mice displayed 16 specific dysregulations, mostly up-regulations in the PVN and PFC, along with up- and down-regulations in other structures. *Shank3* KO mice showed a down-regulation of *Cd38* in the NAC and PFC, as well as shared down-regulations of *Ctsa* and *Rnpep* in the NAC with *Fmr1* KO mice. Notably, none of these mRNAs were affected in *Oprm1* KO mice.

To identify the most discriminant mRNAs for model stratification, we applied stringent criteria, requiring a minimum of 6 total and 4 unique dysregulations across three structures and three models. Within the OT family, only *Oxt, Avp* and *Oxtr* met these criteria. *Oxt* exhibited 7 dysregulations across all five structures and three models, while *Avp* and *Oxtr* showed 8 and 6 dysregulations, respectively, spanning four structures and four models (**Figures S9-S11**).

In conclusion, within the OT family, *Oxt*, *Avp* and *Oxtr* emerge as the most discriminant markers among the four models. These findings highlight the potential utility of these markers in effectively stratifying autism mouse models based on their unique molecular signatures, rather than revealing shared mechanisms.

**Widespread dysregulations in social plasticity across mouse models**

Akin to the OT family, we investigated dysregulations in social plasticity by testing *Arc*, *Egr1, Fos*, *Foxp1* and *Homer1a* — markers identified in WT mice — across the four models following SI unknown at 45 minutes and under basal conditions*.* Our results revealed extensive dysregulations in IEGs across all five structures and mouse models (**Figures 4**, **S12-S13**, **Table S5**). We identified 6 dysregulations in all four models (SON: *Fos*, *Foxp1* and *Homer1a*; PVN, *Egr1* and *Homer1a*; CPU: *Foxp1*), as well as 4 additional dysregulations in three models (SON: *Egr1*; PFC: *Homer1a*; NAC: *Fos*; CPU: *Homer1a*). However, the underlying causes of these dysregulations varied among the mouse models. In the PVN and SON, IEGs remained at basal levels following SI unknown in *Oprm1* KO mice, while in the other models, IEGs were already induced under basal conditions compared to WT mice. Conversely, in the CPU and PFC, the opposite trend was observed.

Applying the same criteria as the OT family, we identified *Egr1*, *Fos, Foxp1* and *Homer1a*, with *Homer1a* standing out with 16 total and 7 unique dysregulations across all five structures and all four models. Surprisingly, in contrast to the findings in WT animals, *Arc* was only dysregulated in the PFC of *Shank3* KO mice (**Figure S12**).

In conclusion, our study unveils widespread defects in social plasticity across the models. *Egr1*, *Fos, Foxp1* and *Homer1a* emerge as the most dysregulated IEGs in the four mouse models, providing valuable markers for model stratification.

**Molecular stratification of mouse models using *Egr1*, *Foxp1, Homer1a*, *Oxt* and *Oxtr***

Integration of qPCR data in the four ASD mouse models following SI unknown aimed to verify connections between our mRNA candidates and specific behavioral parameters (**Figure S14**). Component 1 of this analysis unveiled a positive association (SON: *Foxp1*) and negative associations (CPU and PFC: *Fos* and *Homer1a*; PFC: *Foxp1*) with the time spent and mean duration of nose contacts (**Figure S14A**). These associations were primarily driven by isolated mice, aligning with the observed increase in nose contacts within this specific cohort (**Figure S14C**). Conversely, positive associations were found between *Fos* and *Homer1a* in the PFC and *Fos* in the CPU, with the number of rearing events and *Oprm1* KO mice, indicating a potential opposition between spatial exploration and social interaction. Additionally, grooming behavior was associated with *Shank3* KO mice, thus validating our analysis (**Figure S14B**). This association showed a positive correlation with *Homer1a* in the PVN and a negative correlation with *Oxtr* in the PFC. All together, these findings confirm the roles of *Fos, Foxp1, Homer1a* and *Oxtr* as molecular markers linked to core features of autism in mice.

In the last analysis, we cross-referenced data from WT mice and the four mouse models, identifying *Egr1*, *Foxp1, Homer1a*, *Oxt* and *Oxtr*, as robust molecular markers, while excluding *Fos* and *Avp*. Notably, *Fos* and *Avp* exhibited less regulation compared to these five markers. In addition, the high correlation between *Avp* and *Oxt* could bias the model towards their shared dysregulations. Employing these five molecular markers, we conducted a proof-of-concept for potential stratification among the four mouse models (**Figure 5**). The linear discriminant analysis (LDA) integrated molecular data from these markers across the five brain structures and the four models (excluding WT) under basal conditions and following SI unknown (**Figure 5A-B**). The analysis unveiled distinct classifications, identifying *Oprm1* KO (LD1) and isolated mice (LD2) as different models. Although *Shank3* and *Fmr1* KO mice clustered together, they exhibited individual characteristics (LD3). Among the markers, levels of *Oxt* (LD1-3) and *Homer1a* in the CPU (LD1), as well as *Homer1a* in the NAC (LD2-3) and SON (LD1) exerted the most significant influence on the stratification, followed by *Oxtr* across the structures.

To challenge our stratification, we tested *Arc* KO mice, previously documented to manifest social interaction impairments [58, 73]. Based on the five markers, LDA predicted that *Arc* KO mice would be the closest to *Fmr1* KO mice, followed by *Shank3* KO mice (**Figure 5A**). Indeed, *Arc* KO mice displayed an intermediate phenotype, showing social interaction impairments coupled with increased self-grooming (**Figure S15**). Both molecular and behavioral data confirmed *Arc* KO mice as a valid mouse model with ASD features. Additionally, we employed our stratification to predict the potential responsiveness of subgroups (e.g., *Fmr1* and *Shank3* KO mice) to treatment administration (**Figure 5C**). Solely considering the *Oxt* and *Oxtr* markers, we pinpointed cluster 1, consisting of *Fmr1* KO and *Shank3* KO mice with low levels of *Oxt* and *Oxtr*, suggesting a potential positive response to oxytocin treatment in this subgroup.

In conclusion, this study showed the first successful proof-of-concept for the stratification of four mouse models using five molecular markers, providing a potential framework for stratification of individuals with autism.

## DISCUSSION

Our findings elucidate distinct dynamics in the expression patterns of the two mRNA families across five brain structures in response to two types of social interactions in WT mice. Notably, SI mate elicited a sustained pattern of expression, unlike NSI object, which displayed a rapid and transient increase, indicating a response to the novelty of the test environment. SI unknown exhibited a unique pattern within 45 minutes, setting it apart from the other two stimuli. Furthermore, our results suggest a distinct molecular mechanism of activation for each social interaction in these five structures. SI unknown predominantly impacts the OT family in the CPU, NAC and PVN, as well as the IEG family in the PVN, SON and PFC. Conversely, SI mate primarily influences the OT family in the PVN, along with the IEG family in the PFC. While previous studies have highlighted the role of the CPU in facilitating mate interactions as a habitual behavior, and the NAC in orchestrating responses to SI unknown [74], our results reveal a more intricate and nuanced molecular mechanism across these structures.

Surprisingly, our findings unveiled a remarkably high positive correlation between *Oxt* and *Avp* mRNA levels in both WT mice and ASD mouse models. *Oxt* and *Avp*, being paralog genes, are located only 11 kbp apart from each other (4 kbp in humans). Despite the identification of specific elements in this intergenic region *in vitro* that potentially promote neuron specificity [75], our results suggest that both mRNAs undergo similar regulation, and common regulatory elements may have been evolutionarily conserved. The robust correlation observed between *Oxt* and *Avp* mRNAs emphasizes the importance of considering both peptides in research studies.

Within the OT system, our results revealed distinct molecular dysregulations in each mouse model. Particularly, levels of *Oxt* and *Avp* mRNAs, as well as *Oxtr* across the structures delineated the four mouse models of ASD. Noteworthy, *Shank3* KO mice exhibited *Cd38* downregulation in the PFC and NAC. Given its association with social memory and recognition impairments [76], and the social novelty phenotype displayed by *Shank3* KO mice, our results position *Cd38* as a potential marker of social memory-related conditions. In contrast to the OT family, our results underscore widespread IEG dysregulations across brain structures in all four models, confirming synaptic plasticity impairment as a major hallmark of ASD. Notably, *Egr1*, *Foxp1* and *Homer1a* emerge as the most robust markers of social interactions in WT animals and the most dysregulated IEGs in the four models. These findings align with their previous association with ASD in both individuals with autism and mouse models [52, 77–81].

Our results revealed robust impairments in social interaction with an unknown conspecific in *Fmr1* and *Shank3* KO mice, accompanied by pronounced motor stereotypies in *Shank3* KO mice, aligning with previous reports [82, 83]. Conversely, *Oprm1* KO and chronically isolated mice did not exhibit the expected social impairments [69, 71]. Their phenotype may have been influenced by the experimenter's sex, predominantly females in this study [84], as well as by experimental conditions designed to reduce animal stress. Indeed, a slight increase in light induced social impairments in *Oprm1* KO mice. Recently, it has been proposed that housing conditions and the duration of social isolation may interfere with social impairments previously observed in chronically isolated animals [85, 86]. Interestingly, none of the mouse models exhibited impaired social interaction with a cage mate, aligning with reports suggesting that interactions within the family environment were more manageable for children with ASD [87]. Notably, the stratification of the models using the five molecular markers — *Egr1*, *Foxp1*, *Homer1a*, *Oxt*, and *Oxtr* — consistently aligned with behavioral data. This analysis accurately predicted the molecular underpinnings of the observed behavioral differences between *Oprm1* KO and isolated mice and the other two models, as well as *Arc*

KO mice. Despite their distinct profiles, it unveiled shared dysregulations in both *Fmr1* and *Shank3* KO mice, providing valuable insights into potential links with their respective social phenotypes and responsiveness to treatments.

In conclusion, our study offers valuable insights into the dynamics of OT and IEG mRNAs in WT animals and their dysregulations in mouse models of ASD across five structures within the social and emotional neuronal circuit. Our findings not only enable the stratification of the four mouse models, but also allow for the identification of subgroups within these models. The stratification of other mouse models of ASD, such as *Cntnap2*, *Magel2* or *Oxtr* KO mice [88–90], and different species, including *Shank3* KO rats [91], would enhance the predictive value of the five molecular markers in individuals affected by autism. Exploring additional structures within this circuit, such as the amygdala and lateral septum, could further contribute to the stratification of models. Nevertheless, future research employing omic approaches beyond the oxytocin and synaptic plasticity families is essential to identify more molecular markers for social interaction and potentially uncover novel therapeutic targets. This study represents the first proof-of-concept for the potential stratification of individuals with autism using mouse models of ASD, with the aim of improving the success of clinical trials and personalized treatment for ASD. Additionally, our work holds promise for precise and faster ASD diagnostics, complementing behavioral screenings (Diagnostic and Statistical Manual of Mental Disorders 5). The development of these molecular markers into screening assays, utilizing accessible fluids like blood, could streamline diagnostics, as demonstrated by the feasibility of detecting elevated Arc proteins or decrease OT peptide levels in the blood of boys with ASD [26, 50].

**ACKNOWLEDGEMENTS**

Mouse breeding and care was performed at INRAE Animal Physiology Facility (https://doi.org/10.15454/1.5573896321728955E12). This work has benefited from the facilities and expertise of the "Informatique Scientifique Locale et Analyses de Données" (ISLANDe) of the UMR PRC, INRAE. We used ChatGPT, developed by OpenAI, for assistance with English editing. We thank Dr E. Valjent, Dr R. Yvinec, Dr P. Crepieux, Dr P. Chamero and Dr J. Collet for their advice on the manuscript.

This project has received funding from the European Research Council (ERC) under the European Union's Horizon 2020 research and innovation programme (grant agreement No. 851231). This work was supported by the INRAE "SOCIALOME" project (PAF_29). LPP, CG and AD acknowledge the LabEx MabImprove (grant ANR-10-LABX-53-01) for the financial support of CG and AD PhD's co-fund.

**Data availability statement**

All the data are available in **supplementary Tables**. Movies of behavioral experiments are available upon request.

**Conflict of interest**

The authors declare no conflict of interest.

**Author contributions**

CG, AD, OV, LD, EP and LP designed and performed the experiments; CG, AD, OV, LD, EP and LP contributed to the data collection; CG, AD, OV, LD, GL and LP contributed to the interpretation of data; GL performed all statistical analysis, data modeling and integration; CG, GL and LP wrote the original drafts; CG, AD, OV, LD, EP, GL and LP reviewed and edited the manuscript; LP contributed to the funding acquisition, project conceptualization and supervision.

**Abbreviations**

ASD autism spectrum disorders

AVP vasopressin

CPU caudate putamen

HCPC hierarchical clustering on the principal components of PCA

IEG immediate early gene

KO knockout

LDA linear discriminant analysis

NAC nucleus accumbens

NSI non-social interaction

OT oxytocin

PCA Principal Component Analysis

PFC prefrontal cortex

PVN paraventricular nucleus of the hypothalamus

SI social interaction

SON supraoptic nucleus of the hypothalamus

WT wildtype

**Supplementary information** is available at MP’s website.

## FIGURE LEGENDS

### Figure 1 All four models exhibited distinct behavioral features along the autism spectrum

In the reciprocal social interaction tests, the cumulative time in nose contacts with a genotype-, sex- and age-matched unknown mouse (**A**) is reduced in *Fmr1* KO mice (green), but not in *Oprm1* (blue), *Shank3* KO (yellow) and isolated (burgundy) mice, compared to WT animals (gray). In motor stereotypies, the time spent in self-grooming (**B**) is increased in *Shank3* KO mice and reduced in *Oprm1* and *Fmr1* KO animals, while the number of head shakes (**C**) is increased in *Shank3* KO mice and reduced in isolated mice, compared to WT mice. In the sociability (**D**), the social novelty (**E**) or mate preference (**F**) phases of the three-chambered test, the cumulative time in nose contacts of *Oprm1, Fmr1, Shank3* KO and isolated mice are compared to WT animals. In the sociability phase, *Fmr1* KO mice exhibit reduced time spent in nose contact with an unknown WT mouse (chamber A) compared to WT mice, but not with the object (chamber B). In the social novelty phase, *Shank3* KO mice display no preference for the novel mice (unknown WT mouse, chamber B) over the familiar mouse (already explored for 10 minutes in the precedent phase, chamber A). *Oprm1* KO mice show a lack of preference for the cage mate (chamber A) over the familiar mouse (already explored for 10 minutes in the precedent phase, chamber B). Data are presented as individual data, mean ± sd (**Table S3**). Kruskal-Wallis tests followed by Dunn *post hoc* tests were performed with stars as line effect and hash as chamber effect (p = adjusted p-value with Benjamini-Hochberg correction). * or # $p < 0.05$, ** or ## $p < 0.01$, *** or ### $p < 0.001$, **** or #### $p < 0.0001$.

**Figure 2** ***Arc*, *Avp, Avpr1a, Egr1*, *Fos*, *Foxp1, Homer1a, Oxt* and *Oxtr* are the most discriminant mRNAs in response to social stimuli in WT mice**

Levels of *Avp, Avpr1a, Oxt* and *Oxtr*, as well as *Arc, Egr1*, *Fos*, *Foxp1,* and *Homer1a,* among the OT (blue) and IEG (orange) families under basal conditions and 45 minutes, 2 or 6 hours following interactions with an unknown mouse (SI unknown), a cage mate (SI mate) or an object (NSI object) in WT animals are represented in one heatmap. Blue and red indicate low and high expression levels, respectively. Hierarchical clustering identifies three clusters among the different interactions and time points: one cluster with SI unknown at 45 minutes, one with SI mate at 45 minutes, 2 hours and NSI object at 45 min, and one cluster with the other conditions. mRNAs and structures were separated in 7 clusters (1 to 7 from left to right). Clusters 1 and 2 mainly discriminate mRNAs induced by SI unknown at 45 minutes in the OT family in the NAC, CPU and PFC and IEGs in the PVN and SON, respectively. Clusters 4 and 7 highlight mRNAs that are not induced by SI unknown at 45 minutes compared to NSI object and SI mate, while cluster 6 shows the mRNAs that are induced by both social stimuli. Clusters 3 and 5 represent mRNAs induced by SI mate and NSI object or NSI object, respectively. All data, mean and statistical values are represented in **Figures S5-S7** and detailed in **Table S4**. CPU, caudate putamen; NAC, nucleus accumbens; PFC, prefrontal cortex; PVN, paraventricular nucleus; SON, supraoptic nucleus.

**Figure 3** **Dysregulations in the OT family are model-specific**

In the CPU (**A**), except *Oxtr* mRNAs in isolated mice, levels of *Avp* and *Oxtr* genes were not significantly induced 45 minutes after social interaction with an unknown mouse (SI unknown, dashed lines) over basal conditions (solid lines) in *Oprm1* (blue), *Shank3* KO (yellow), *Fmr1* KO (green) and isolated (burgundy) mice, compared to WT mice (gray). Except for these shared dysregulations, dysregulations are rather specific to a model. In the PVN (**B**), *Fmr1* KO mice exhibit reduced levels of *Oxt, Avp, Oxtr* and *Avpr1a* mRNAs under basal conditions, compared to WT mice. In the PFC (**C**), *Avp, Oxt* and *Oxtr* mRNAs are up-regulated in isolated mice (burgundy), compared to WT mice. Data are presented as individual data, mean ± sd (**Table S5**). Linear models followed by *post hoc* tests based on estimated marginal means were performed with stars as line effect and hash as basal *vs* SI unknown 45 minutes effect (p = adjusted p-value with Tukey correction). * or #p < 0.05, ** or ##p < 0.01, *** or ###p < 0.001. CPU, caudate putamen; PFC, prefrontal cortex; PVN, paraventricular nucleus.

**Figure 4 Among the IEGs, *Egr1*, *Fos*, *Foxp1* and *Homer1a* are not induced across the four mouse models**

In the SON (**A**), PVN (**B**) and CPU (**C**), *Egr1, Fos, Foxp1* and *Homer1a* levels were not induced 45 minutes after social interaction with an unknown mouse (SI unknown, dashed lines) over basal conditions (solid lines) across *Oprm1* (blue), *Shank3* KO (yellow), *Fmr1* KO (green) and isolated (burgundy) mice, compared to WT mice (gray)**,** except *Egr1* for *Fmr1* KO mice. Notably, in the SON, *Fos, Foxp1* and *Homer1a* were not induced for different reasons in these models. While they remained at basal levels following SI unknown in *Oprm1* KO mice, they were already induced under basal conditions in the other three models. Data are presented as individual data, mean ± sd (**Table S5**). Linear models followed by *post hoc* tests based on estimated marginal means were performed with stars as line effect and hash as basal *vs* SI unknown 45 minutes effect (p = adjusted p-value with Tukey correction). * or #p < 0.05, ** or

##p < 0.01, *** or ###p < 0.001, **** or ####p < 0.001. CPU, caudate putamen; PVN, paraventricular nucleus; SON, supraoptic nucleus.

**Figure 5 Stratification of the four models using the 5 molecular markers *Egr1, Foxp1, Homer1a, Oxt* and *Oxtr***

Linear discriminant analysis (LDA; **A**) revealed that *Oprm1* (blue, LD1 component) and isolated (burgundy; LD2 component) mice clustered apart from the other two models. LD3 component discriminates between *Shank3* KO (yellow) and *Fmr1* KO (green) mice. While challenged with *Arc* KO mice (black), the stratification predicted that this new mouse model was similar to *Fmr1* KO mice (class membership probability of 6), followed by *Shank3* KO mice (probability of 2), and *Oprm1* KO and isolated mice (probability of 1). The heatmap with LDA coefficients (**B**) indicates that the five molecular markers, *Egr1, Foxp1, Homer1a, Oxt* and *Oxtr,* contributed to the stratification of the models, with *Oxt* and *Homer1a* in the CPU and *Homer1a* in the NAC and SON contributing the most. Hierarchical Clustering on Principal Components (HCPC; **C**) revealed three subgroups among *Shank3* KO and *Fmr1* KO mice (left panel). Cluster 1, which contains *Shank3* KO and *Fmr1* KO mice, displays lower levels of *Oxt* and *Oxtr* in the CPU, NAC and PFC, as well as increased levels of *Oxtr* in the PVN. Blue, red and white indicate low, high levels or no significant contribution, respectively. PFC, prefrontal cortex; NAC, nucleus accumbens; CPU, caudate putamen; PVN, paraventricular nucleus; SON, supraoptic nucleus.

**Figure 1** **All four models exhibited distinct behavioral features along the autism spectrum**

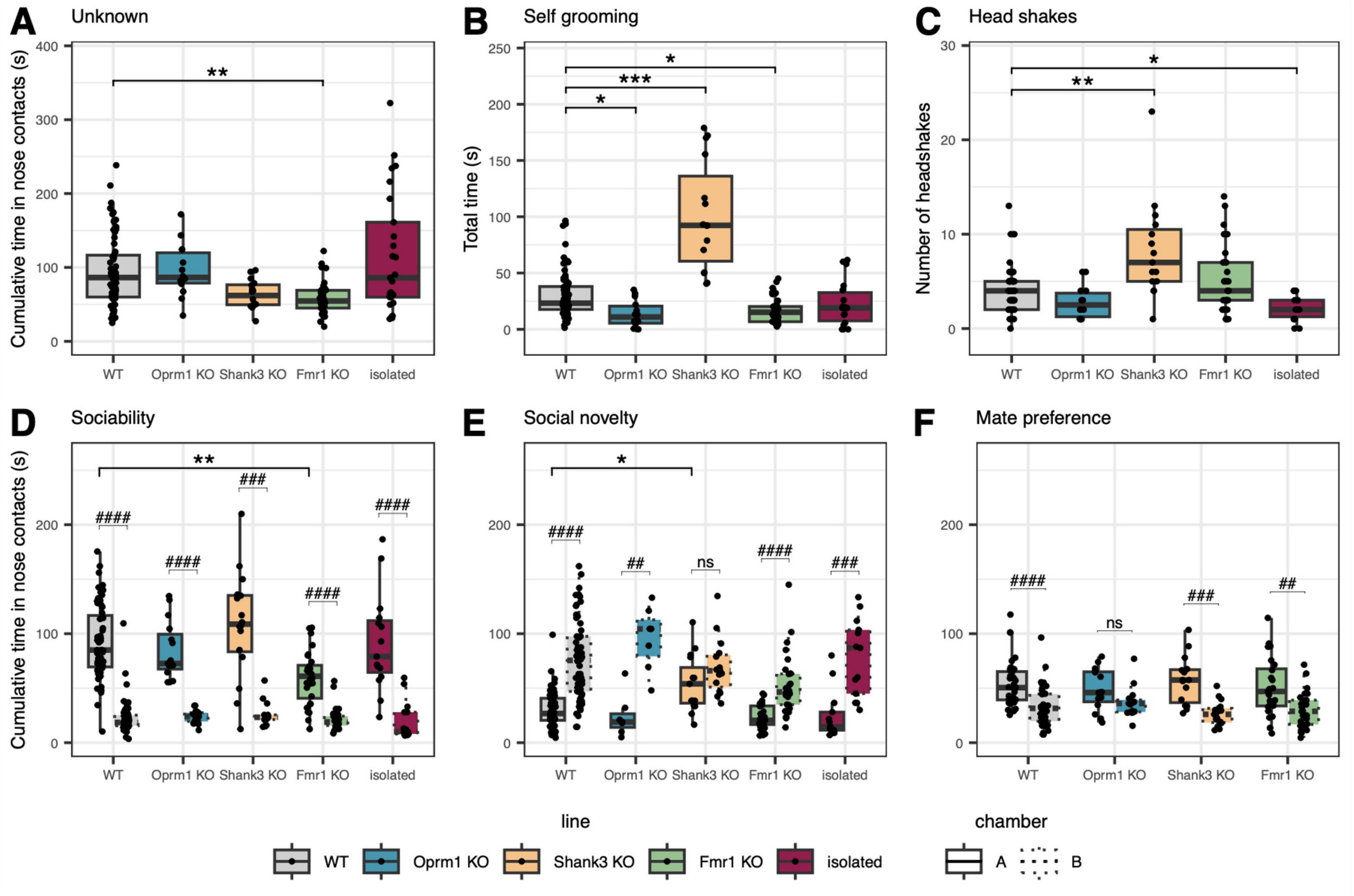

**Figure 2** ***Arc*, *Avp*, *Avpr1a*, *Egr1*, *Fos*, *Foxp1*, *Homer1a*, *Oxt* and *Oxtr* are the most discriminant mRNAs in response to social stimuli in WT mice**

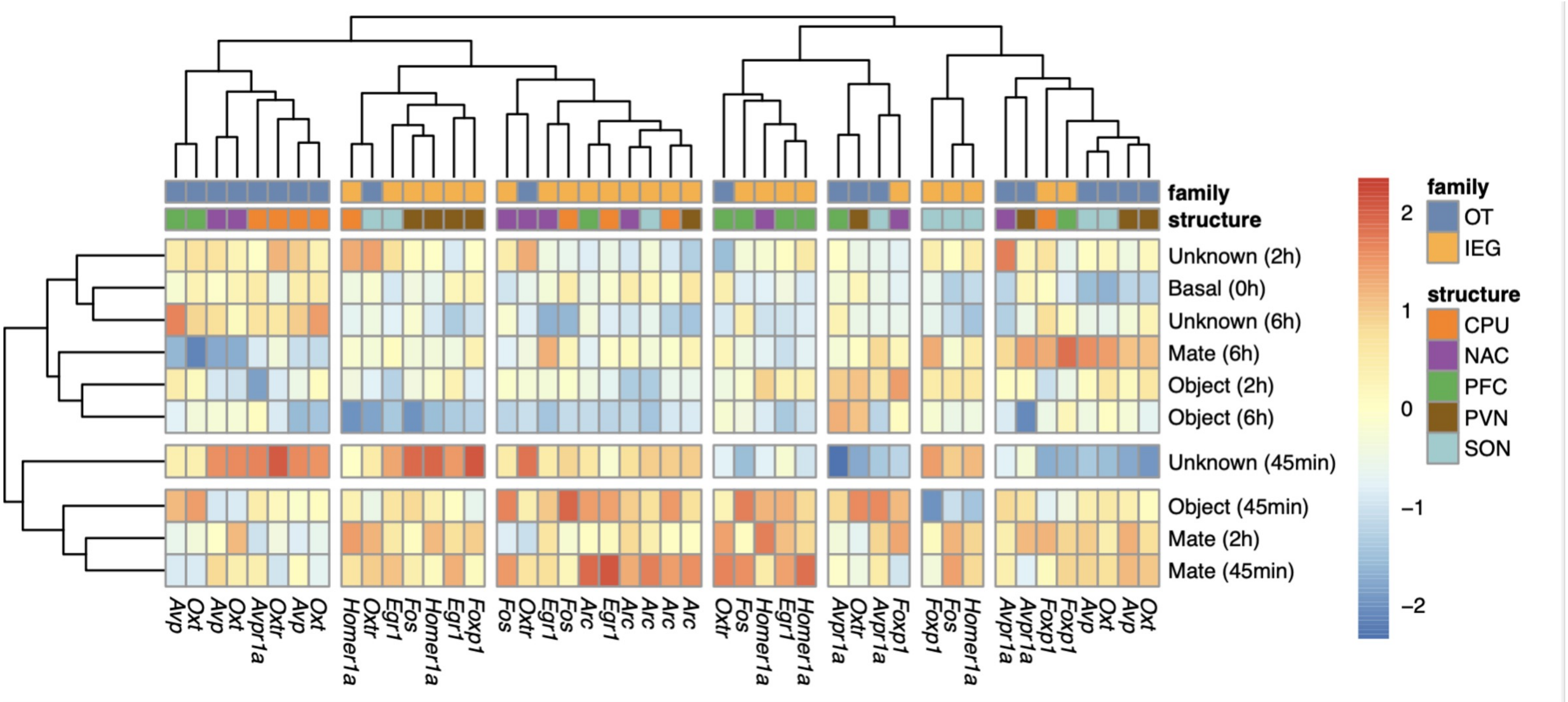

**Figure 3** **Dysregulations in the OT family are model-specific**

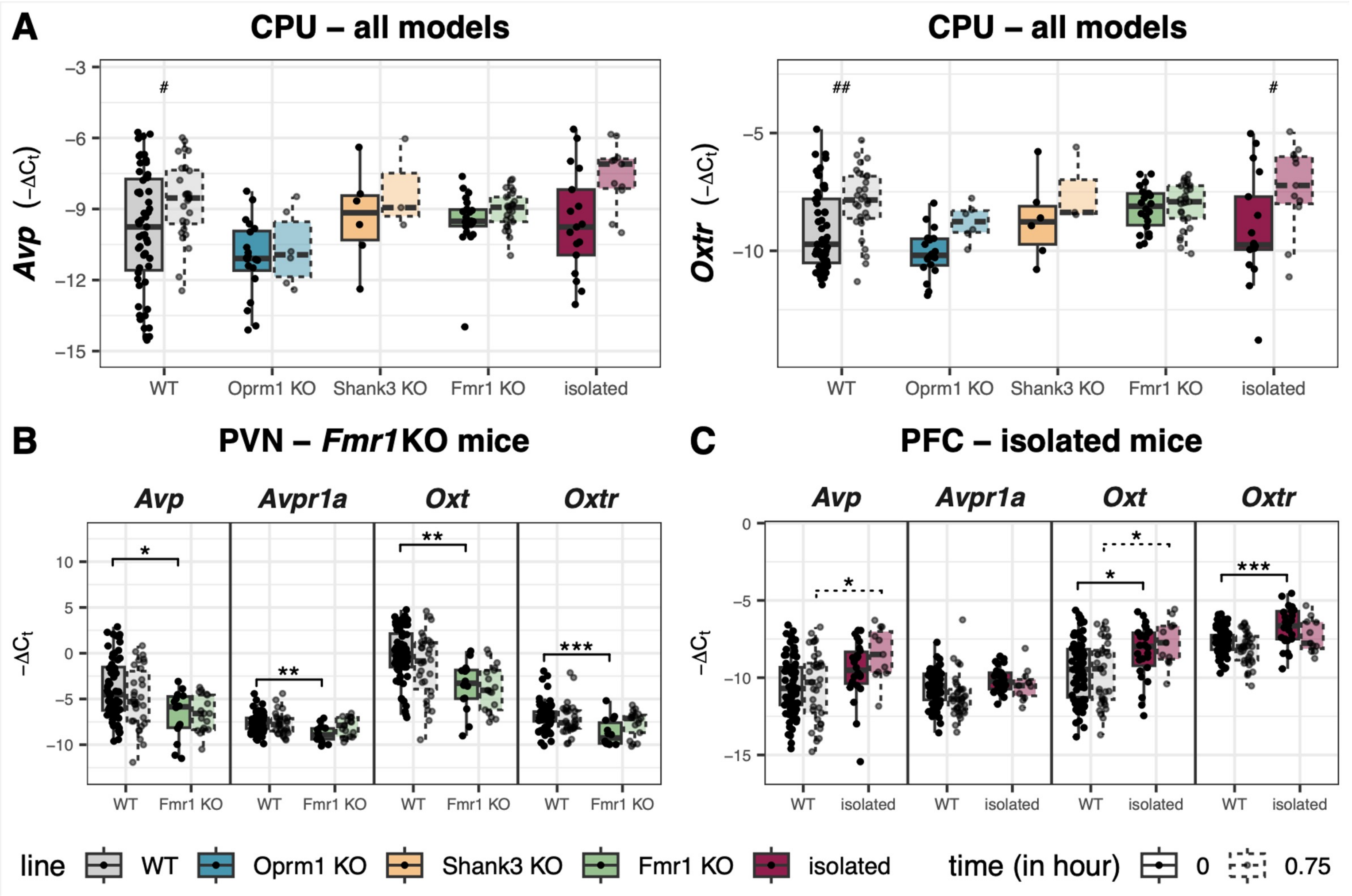

**Figure 4** Among the IEGs, *Egr1*, *Fos*, *Foxp1* and *Homer1a* are not induced across the four mouse models

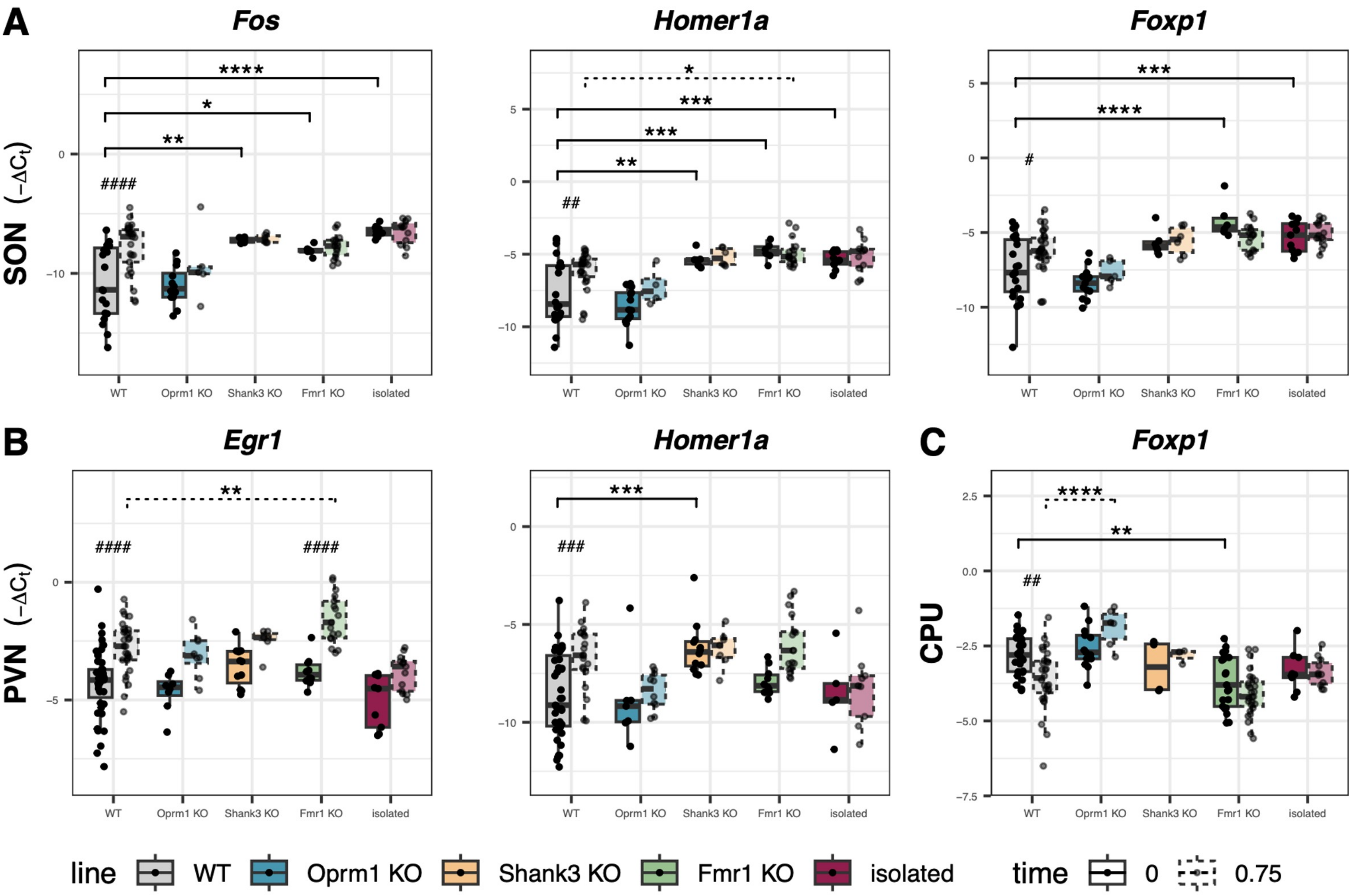

**Figure 5 Stratification of the four models using the 5 molecular markers *Egr1*, *Foxp1, Homer1a, Oxt* and *Oxtr***

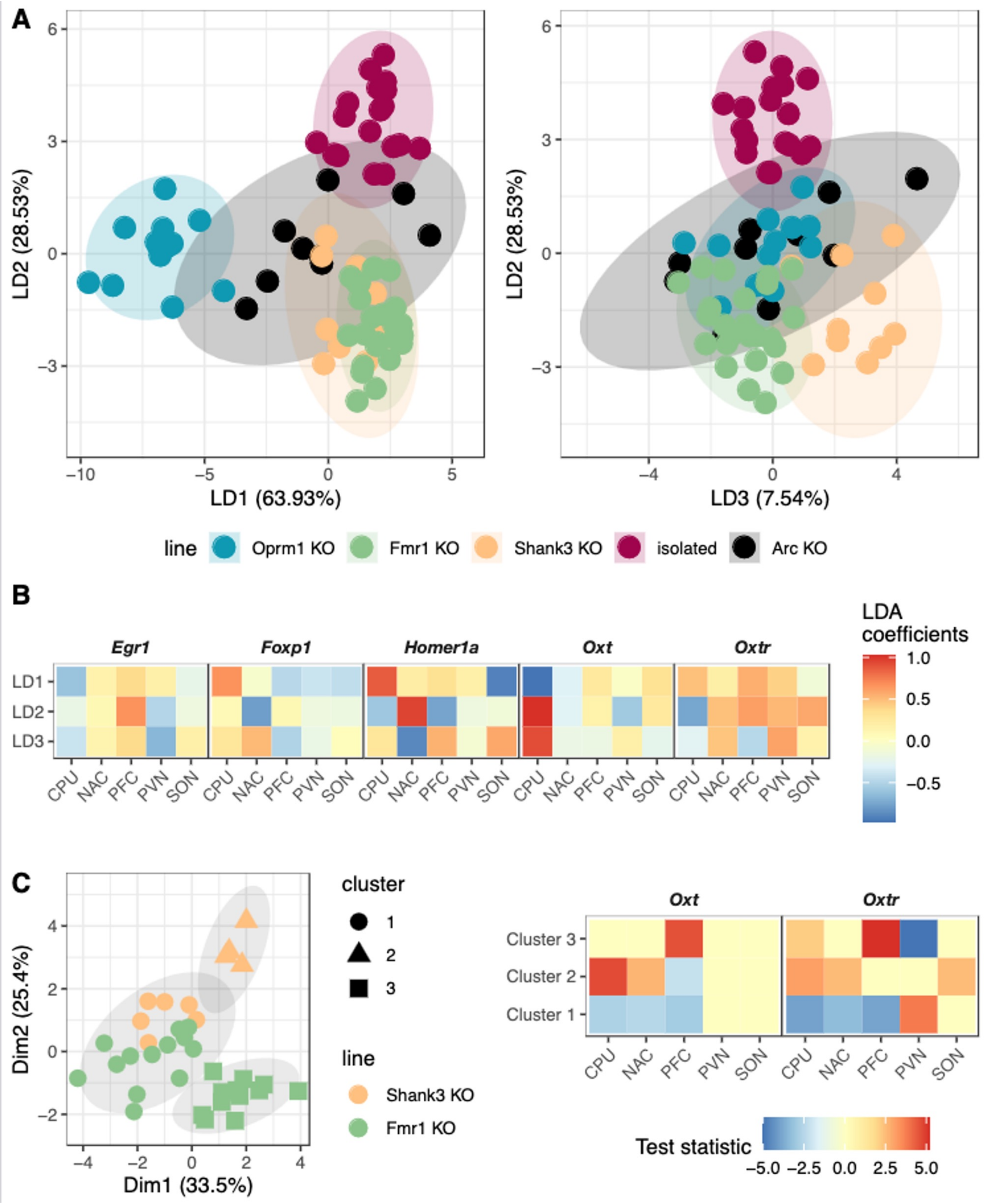

# SUPPLEMENTARY INFORMATION

# SUPPLEMENTARY MATERIALS AND METHODS

## Behavior experiments

Following heterozygous breeding scheme, 2 independent cohorts for each mouse lines were generated from a minimum of three different homozygous non-inbred couples, maintained on a mixed (exactly 50%-50%) C57BL/6J;129S2 background in the same breeding room of the animal facility. All behavioral tests were carried out, when possible, on mornings to avoid any circadian cycle effect, with one behavioral test per day in a dedicated quiet room with controlled temperature and humidity, with a dim light intensity of 15 lux (40 lux for motor stereotypies). Behavioral equipment was built and acquired according to standard size and height in literature and manufacturer for robustness and reproducibility (see dimensions in each subsection). All floors were covered with an aluminum foil coated with a textured, non-reflective gray epoxy neutral paint to favor normal locomotion, well-being and reduced anxious-like behaviors. Each type of behavior was assessed in at least two behavioral paradigms. The experimental design was as followed, reciprocal social interaction tests between two sex-, genotype- and age-matched unknown mice, named 'SI unknown' at Day 0 (D0), or cage mate mice, named 'SI mate' at D1, interaction with an object as a non-social control interaction, named 'NSI object' (D3), Y-maze (D4) and two resting days followed by the three-chamber (D7-8) and motor stereotypies tests (D9).

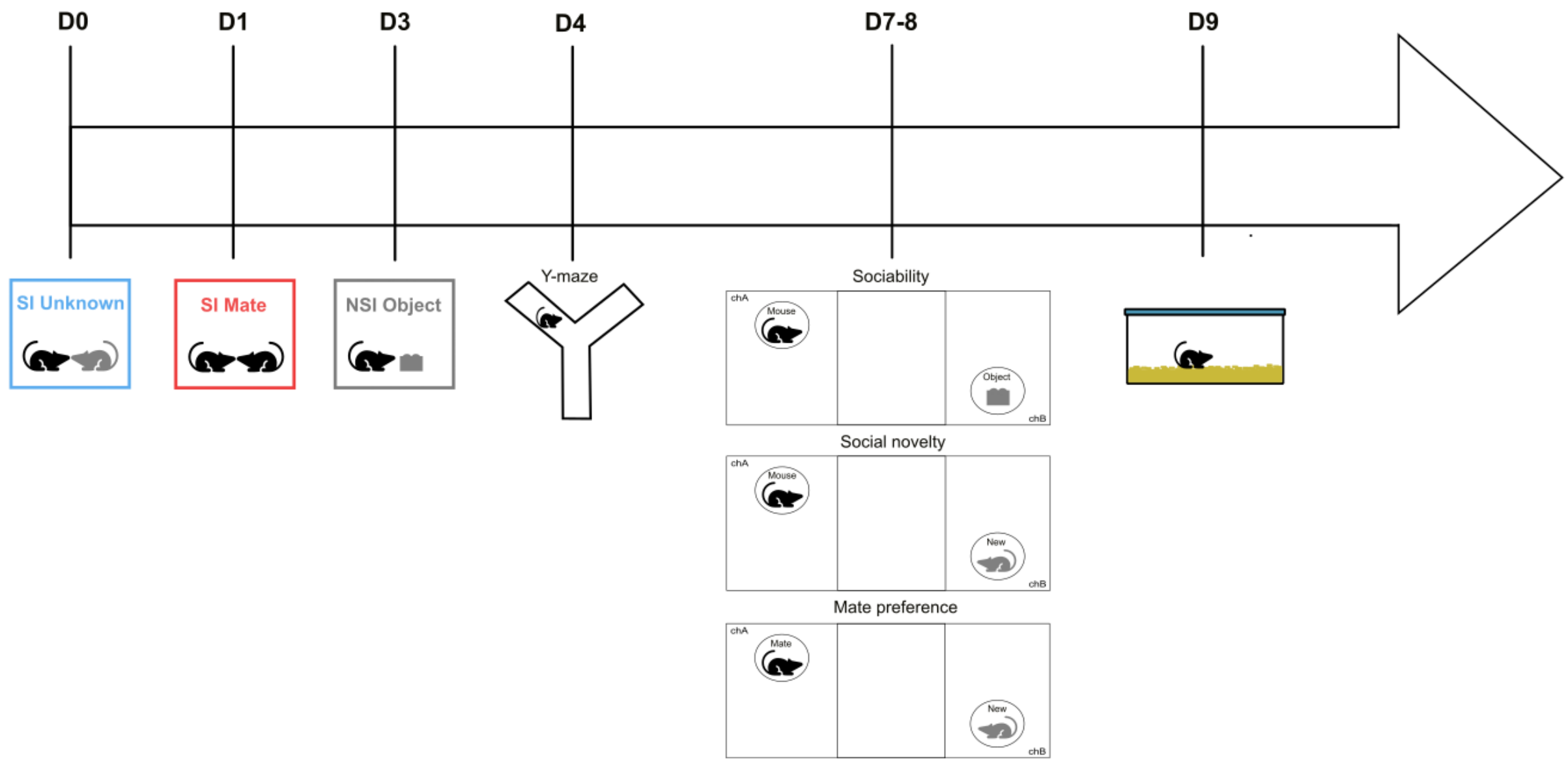

All the videos were recorded from above using an USB black and white camera with 2.8-12 mm varifocal optic and ANY-maze software, except for the motor stereotypies test that was recorded from the side using a Sony HD FDR-AX33 4K Camescope. Mice locomotion, interaction (cylinders, object) and anxious-like behaviors (e.g., time spent in the center or border zones of the open field arena) in the habituation phase of three-chambered test, Y-maze and object interaction tests were analyzed using the automatic animal tracking ANY-maze software (Stoelting, Ireland). Motor stereotypies and social interactions (SI unknown, SI mate and sociability, social novelty and mate preference phases in three-chambered test) were manually scored by one trained experimenter blind to genotype and/or housing conditions using the Behavioral Observation Research Interactive Software (BORIS) (Friard and Gamba, 2016). The automatic ANY-maze software was configured to detect animal immobility using a 95% threshold sensitivity, animal entries in compartment or arms using 80% of the animal body (e.g., when 4 paws of the animals entered the receptive chamber or arms in the three-chambered and Y-maze tests), and the mouse head for object interaction. Criteria to exclude animals of behavioral tests were over 30% of the time spent immobile and/or lack of exploration of all arms or compartments. In all tests, “nose contacts” is defined as nose-to-nose, nose-to-body or nose-to-anogenital region contacts.

## Object interaction test

Mice were placed 10 minutes at 15 lux in an arena (Ugo Basile, Italy; 100 x 100 cm) divided by 4 dark gray opaque partitions and walls in 4 open fields (46 x 46 cm) over a granular and non-reflective gray floor that provides an optimal contact for the mice. In this object interaction test, a Lego Duplo toy was taped to the center of the arena using adhesive pasta. The total time, mean duration of nose contact with the object, number of visits, first latency, immobility, total distance traveled, cumulative time spent and entrance in the central squared zone (40 x 40 cm) and in the border wall zones were measured automatically using the ANY-maze software.

## Reciprocal social interaction test

Each experimental mouse met an age-, sex- and genotype-matched unknown conspecific or a cage mate in the open field. The total amount of time, the number and mean duration spent in nose contact, self-grooming, grooming within 5 seconds following social contact, paw contacts, attacks, following, huddling, immobility and the number of rearing, vertical jump and circling events were measured by a blinded experimenter using BORIS software.

## Three-chambered test

The three-chambered social apparatus (Ugo Basile, Italy) (PMID: 15666335) displays three equal compartments (20 x 40 x 22 cm), with 3 dark gray opaque and one transparent (hidden by gray walls from the other apparatus) PVC walls and gray granular and non-reflective floor optimal for animal welfare and tracking. Chambers are separated by transparent internal partitions with two sliding doors (5 x 8 cm) that allow transitions between chambers. Two-cylinder grid cages (7 cm diameter x 15 cm, 0.3 cm diameter-rods spaced 0.7 cm apart with gray floor and transparent lids) were located within the opposite chambers. 7-8-week-old wild-type female and male mice, namely 'interactor' were trained in the grid cages for at least two days prior to the test (>20 min/day) to avoid any spatial or anxious-like bias and were randomly assigned to the grid cages located in the opposite chambers (up/left or down/right). On the testing day, the experimental animal was placed in the middle chamber and allowed to explore the whole apparatus for 10 minutes at dim light (15 lux, habituation phase). Thereafter, three consecutive phases of 10 minutes were performed: the sociability phase (sex-matched unknown interactor 1 *vs* Lego Duplo toy), the social novelty phase (interactor 1 *vs* a novel sex-matched unknown interactor 2 replacing the toy) and the mate preference phase (interactor 2 *vs* a cage mate replacing interactor 1). For each phase, the experimental mouse was placed back to the middle chamber and the phase started when interactors or objects were in the grid cages and sliding doors were removed. An animal entry was valid when a mouse entered its 4 paws in a chamber. The total amount of time, number and mean duration of compartment entries, nose contacts to interactor, object or empty grid cages, the total distance traveled, self-grooming episodes and immobility were measured by ANY-maze (habituation) or an experimenter blind to genotype using BORIS (sociability, social novelty and mate preference). The time spent by experimental animals exploring the top half of the grid cage was not considered as interaction unless the stimulus animal was also at the top of the grid cage.

## Motor stereotypies

Experimental animals were recorded individually in a clear type I cage (33,1 x 15,9 x 13,2 cm) filled with a thick layer of 3-4 cm litter for 10 minutes at 40 lux. The total time, number and mean duration of self-grooming and digging for repetitive behaviors, the number of vertical jumping, circling, head shakes and scratching episodes for stereotyped behaviors, as well as the total time spent immobile and the number of rearing events were manually scored a posteriori by an experimenter blind to genotype using BORIS.

## Spontaneous alternations in the Y-maze test

Y-maze apparatus (Ugo Basile, Italy) displays dark gray wall PVC arms of standard size (35 x 5 x 10 cm) with a metal base painted in non-reflective and granular light gray floor that can be disassembled and closed with the included doors. Each mouse was placed in one arm of the Y-maze and allowed to freely explore the three arms for 5 minutes at 15 lux. The percentage of spontaneous alternation (SPA), alternative arm returns (AAR) and same arm returns (SAR) determined from triplet of arm entries was measured, as well as the total distance traveled, the mean speed, the number of entries (entry of the 4 paws inside an arm), the total time, number and mean duration spent in each arm using automatic ANY-maze software.

## Behavioral tests for quantitative PCR

Two month-old male and female mice were subjected to 10 minutes social interaction with an unknown ("SI unknown") or a cage mate ("SI mate") mouse of the same sex, housing condition and genotype in order to prevent any bias when comparing mate vs unknown interaction, or to object interaction (Lego toy, "NSI object") in the open field arenas. After the tests, mice were placed back to their home cage with their cage mates (except isolated mice, which remain isolated in their home cage) until dissection time. "Control" animals (*e.g.*, animals that did not perform any interaction tests) were moved from the breeding room to the behavioral room as the other experimental mice, which were subjected to social interaction tests, but remained in their home cage with at least one cage-mate until dissection. Mice exposed to 6 hours of acute isolation, as a negative social stimulus, remained isolated in their home cage in the behavioral room where cage mates were removed. At least 8 mice from each sex, condition/genotype and interaction were dissected in the morning for "control a.m.", 45 minutes or 2 hours following one of the 3 interactions or in the afternoon for "control p.m.", 6 hours of acute isolation and 6 hours following an interaction.

# SUPPLEMENTARY FIGURES AND TABLES

## Supplementary Tables

**Table S1 List of primers for qPCR and their validation**

**Table S2 List of materials and reagents**

**Table S3 All raw data, mean and statistics from behavior tests**

**Table S4 All raw data, mean and statistics from qPCR data in WT mice**

**Table S5 All raw data, mean and statistics from qPCR data in mouse models**

## Supplementary Figures

**Figure S1 Only qPCR data required batch correction over independent cohorts**

Principal Component Analysis (PCA) showed a cohort effect when analyzing the whole WT dataset (**A**). Data were corrected for batch effect and presented after correction (**B**). PCA revealed that the five brain structures formed two clusters (**C**), one with the PFC, NAC and CPU and one with the PVN and SON. PCA performed on independent cohorts of WT mice revealed no batch effect in behavior parameters (**D**). Schematic representation of the study design (**E**) with the PVN, SON, PFC, NAC and CPU in WT mice exposed to social interactions with an unknown mouse (SI unknown, blue), a cage mate (SI mate, red) or a non-social interaction with an object (NSI object, gray). CPU, caudate putamen; NAC, nucleus accumbens; PFC, prefrontal cortex; PVN, paraventricular nucleus; SON, supraoptic nucleus.

**Figure S2** ***Fmr1*** **heterozygous females and** ***Oprm1*** **KO exposed to 40 lux display social interaction impairments**

In the reciprocal social interaction and the three-chambered tests, the cumulative time in nose contacts with a genotype-, sex- and age-matched unknown mouse (**A**, **D**) and in the sociability (**B**, **E**), the social novelty (**C**, **F**) phases of *Fmr1* KO (dark green) and heterozygous females (light green; **A**-**C**) and *Oprm1* KO (blue) and WT mice (gray), exposed to 40 (light color) or 15 lux (dark color) light intensity (**D**-**F**) are compared to WT (gray) at 15 lux. Nose contacts with an unknown conspecific are not different in any of the conditions (**A**, **D**). In the sociability phase of the three-chambered test, the time in nose contacts is reduced in *Fmr1* heterozygous females (**B**) and *Oprm1* KO (**E**) exposed to 40 lux, as well as the time in nose contact with the object in WT exposed to 40 lux. In the social novelty phase, *Fmr1* KO mice (**C**) display no preference for the novel mice (unknown WT mouse, chamber B) over the familiar mouse (already explored for 10 minutes in the precedent phase, chamber A) while *Oprm1* KO mice exposed to 40 lux (**E**) exhibit reduced time in nose contacts with the object. Data are presented as individual data, mean ± sd (**Table S3**). Kruskal-Wallis tests followed by Dunn post hoc tests were performed (p= adjusted p-value in all tests) with stars as line effect and hash as chamber effect. * or #p<0.05, ** or ##p<0.01, ###p<0.001, ####p<0.0001.

**Figure S3** ***Shank3*** **KO mice display stereotyped and anxious-like behaviors**

In the motor stereotypy test, the time spent digging (**A**) and the number of rearing events (**B**) are reduced in *Shank3* KO mice, compared to WT. In the Y-maze test, none of the mouse lines shows impaired cognitive flexibility, as evidenced by a difference in the alternation pattern (**C**), or in the locomotion, as measured by the distance traveled in the open field (**D**). *Shank3* KO mice exhibit anxious-like behaviors as they spend more time in the periphery of the arena (**E**) than WT or other mouse models. Data are presented as individual data, mean ± sd (**Table S3**). Kruskal-Wallis tests followed by Dunn post hoc tests were performed (p= adjusted p-value in all tests). *p<0.05, **p<0.01, ***p<0.001, ****p<0.0001.

**Figure S4 Only *Foxp1* and *Arc* mRNAs in the SON of WT animals exhibit a differential pattern in males and females or exposed to 6 hours acute social isolation**

The effect of sex (**A**) and exposure to 6 hours acute isolation in the home cage (**B**) was only significant for *Foxp1* and *Arc* in the SON. *Foxp1* levels were reduced in males compared to females while *Arc* levels were decreased following acute social isolation. Data, expressed as -ΔCt values are presented as individual data, mean ± sd and detailed in **Table S4**. Kruskal-Wallis tests followed by Benjamini-Hochberg correction were performed with *adjusted p-value < 0.05. SON, supraoptic nucleus.

**Figure S5 Dynamic transcriptional profile of oxytocin, vasopressin and their receptor mRNAs in response to interactions**

Transcriptional profiles of *Oxt, Avp, Oxtr, Avpr1a* and *Avpr1b* mRNAs in WT mice are dynamic in response to social interactions with an unknown mouse (SI unknown or U; blue) or a cage mate (SI mate or M; red) or non-social interaction with an object (NSI object or O; gray) across the five structures. *Oxt* and *Avp* responded rapidly with an opposite pattern between SI unknown and SI mate in the CPU and PVN. *Oxtr* and *Avpr1a* mRNAs are dynamic in response to SI unknown, particularly in the CPU, and to a lower extent to SI mate or NSI object. *Avpr1b* mRNAs did not show any significant difference. Lines highlight a significant difference between two time points for each interaction. Letters indicate a significant difference between two interactions at a specific time point (on top of the curves) or overall pattern over time (in bold, right side of the curves). ANOVA were used to select the significant variables followed by post-hoc tests based on estimated marginal means (p= adjusted p-value in all tests). All data, expressed as -Δ*Ct* values, mean and statistical values are detailed in **Table S4**. M, mate; O, object; U, unknown; CPU, caudate putamen; NAC, nucleus accumbens; PFC, prefrontal cortex; PVN, paraventricular nucleus; SON, supraoptic nucleus.

**Figure S6 Dynamic transcriptional profile of IEGs in the CPU, NAC and PFC**

Transcriptional profiles of *Arc, Bdnf, Egr1, Fos, Fosb, Foxp1, Gdnf, Homer1a* and *Jun* show different kinetics across the CPU, NAC and PFC of WT mice at different time points (0, 0.75 or 45 minutes, 2 or 6 hours) in response to social interactions with an unknown mouse (SI unknown or U; blue), a cage mate (SI mate or M; red) or a non-social interaction with an object (NSI object or O; gray). *Arc, Egr1, Fos, Foxp1* and *Homer1a* showed the most significant differences between social and non-social interactions. Lines highlight a significant difference between two time points for each interaction. Letters indicate a significant difference between two interactions at a specific time point (on top of the curves) or overall pattern over time (in bold, right side of the curves). ANOVA were used to select the significant variables followed by post-hoc tests based on estimated marginal means (p= adjusted p-value in all tests). All data, expressed as -Δ*Ct* values, mean and statistical values are detailed in the **Table S4**. PFC, prefrontal cortex; NAC, nucleus accumbens; CPU, caudate putamen.

**Figure S7 Dynamic transcriptional profile of IEGs in the PVN and SON**

Transcriptional profiles of *Arc, Egr1, Fos, Fosb, Foxp1* and *Homer1a* show different kinetics across the PVN and SON of WT mice at different time points (0, 0.75 or 45 minutes, 2 or 6 hours) in response to social interactions with an unknown mouse (SI unknown or U; blue), a cage mate (SI mate or M; red) or a non-social interaction with an object (NSI object or O; gray). Except *Arc* for SI mate*, Egr1, Fos, Foxp1* and *Homer1a* exhibit the most significant differences for SI unknown in these hypothalamic structures. Lines highlight a significant difference between two time points for each interaction. Letters indicate a significant difference between two interactions at a specific time point (on top of the curves) or overall pattern over time (in bold, right side of the curves). ANOVA were used to select the significant variables followed by post-hoc tests based on estimated marginal means (p= adjusted p-value in all tests). All data, expressed as -Δ*Ct* values, mean and statistical values are detailed in the **Table S4**. PVN, paraventricular nucleus; SON, supraoptic nucleus.

**Figure S8 *Oxt* and *Avp* mRNAs are highly correlated**

*Oxt* and *Avp* mRNA levels are highly correlated in WT (**A**) and the four mouse models (**B**), reaching 92% across the NAC, CPU, SON and PFC and 85 and 87%, respectively, in the PVN. Correlation plot between the different mRNAs in WT animals among the OT and IEG families (**C**) reveals in blue and red, low and high levels of correlation, respectively. In addition to *Oxt* and *Avp*, *Homer1a* is the most positively correlated to *Foxp1* and *Egr1* mRNAs, followed by *Fos* with *Jun* and *Homer1a, Bdnf* and *Jun,* and *Oxtr* with *Avpr1b*. Interestingly, *Homer1a*, *Foxp1* and *Egr1* were negatively correlated to *Oxt* and *Avp*. PFC, prefrontal cortex; NAC, nucleus accumbens; CPU, caudate putamen; PVN, paraventricular nucleus; SON, supraoptic nucleus.

**Figure S9 OT family dysregulations in the CPU, NAC and PFC of mouse models**

Dysregulations of *Avp, Avpr1a, Oxt* and *Oxtr* mRNAs within the OT family in *Oprm1* (blue), *Shank3* KO (yellow), *Fmr1* KO (green) and isolated (burgundy) mice under basal conditions (solid lines) or 45 minutes following social interaction with an unknown mouse (SI unknown, dashed lined) in the CPU, NAC and PFC were compared to levels in WT mice. *Avp* mRNA levels were not induced in response to SI unknown across the four mouse models, as well as *Oxtr* levels in *Oprm1*, *Shank3* and *Fmr1* KO mice, compared to WT animals. *Oxt* levels after SI unknown are lower in the NAC of *Fmr1* and *Oprm1* KO mice. *Avp, Oxt* and *Oxtr* mRNAs were up-regulated in the PFC of isolated mice. Data, expressed as -Δ*Ct* values, are presented as individual data, mean ± sd and detailed in **Table S5**. ANOVA followed by post-hoc tests based on estimated marginal means were performed with stars as line effect and hash as basal *vs* SI unknown 45 minutes effect (p= adjusted p-value with Benjamini-Hochberg correction). * or #p<0.05, ** or ##p<0.01, *** or ###p<0.001. PFC, prefrontal cortex; NAC, nucleus accumbens; CPU, caudate putamen.

**Figure S10 OT family dysregulations in the PVN and SON of mouse models**

Dysregulations of *Oxt, Avp, Oxtr, Avpr1a* mRNAs within the OT family in *Oprm1* (blue), *Shank3* KO (yellow), *Fmr1* KO (green) and isolated (burgundy) mice under basal conditions (solid lines) or 45 minutes following social interaction with an unknown mouse (SI unknown, dashed lines) in the PVN and SON were compared to WT mice. All four mRNAs are down-regulated in the PVN of *Fmr1* KO mice, as well as *Oxt* and *Avpr1a* in the SON, compared to WT animals. *Avpr1a* levels after SI unknown are lower in the SON of *Shank3* and *Oprm1* KO mice. Data, expressed as -Δ*Ct* values, are presented as individual data, mean ± sd and detailed in **Table S5**. ANOVA followed by post-hoc tests based on estimated marginal means were performed with stars as line effect and hash as basal *vs* SI unknown 45 minutes effect (p= adjusted p-value with Benjamini-Hochberg correction). * or #p<0.05, ** or ##p<0.01, *** or ###p<0.001. PVN, paraventricular nucleus; SON supraoptic nucleus.

**Figure S11 mRNA levels in the extended OT family under basal conditions in mouse models**

Dysregulations of receptor (*Avpr1b*) and mRNAs involved in OT secretion (*Cd38*), peptide biosynthesis (*Pcsk1, Pcsk2, Pcsk5* and *Cpe*) and degradation (*Lnpep*, *Ctsa* and *Rnpep*) in *Oprm1* (blue), *Shank3* KO (yellow), *Fmr1* KO (green) and isolated (burgundy) mice compared to WT mice are presented under basal conditions. Data are presented as individual data, mean ± sd (**Table S5**). ANOVA was followed by post-hoc tests. *adjusted p-value <0.05, **<0.01, ***<0.001, ****p<0.0001.

**Figure S12 IEG dysregulations in the CPU, NAC and PFC of mouse models**

Dysregulations of *Arc, Egr1, Fos, Foxp1* and *Homer1a* among the IEG family in *Oprm1* (blue), *Shank3* KO (yellow), *Fmr1* KO (green) and isolated (burgundy) mice were compared to WT mice under basal conditions (solid lines) or 45 minutes following social interaction with an unknown mouse (SI unknown, dashed lines) in the CPU, NAC and PFC. Although for distinct reasons, one dysregulation was shared across all four models (CPU: *Foxp1*) and four dysregulations were common between *Fmr1* and *Shank3* KO mice (NAC: *Fos, Homer1a*) or isolated mice (CPU and PFC: *Homer1a*). Nine dysregulations were specific to a model. Only in the PFC, *Arc* was not significantly induced in *Shank3* KO mice, compared to WT mice. Data, expressed in -$\Delta Ct$ values, are presented as individual data, mean ± sd (**Table S5**). ANOVA followed by post-hoc tests based on estimated marginal means were performed with stars as line effect and hash as basal *vs* SI unknown 45 minutes effect (p= adjusted p-value with Benjamini-Hochberg correction). * or #p<0.05, ** or ##p<0.01, *** or ###p<0.001, **** or ####p<0.001. CPU, caudate putamen; NAC, nucleus accumbens; PFC, prefrontal cortex.

**Figure S13 IEG dysregulations in the PVN and SON of mouse models**

Dysregulations of *Arc, Egr1, Fos, Foxp1* and *Homer1a* among the IEG family in *Oprm1* (blue), *Shank3* KO (yellow), *Fmr1* KO (green) and isolated (burgundy) mice were compared to WT mice under basal conditions (solid lines) or 45 minutes following social interaction with an unknown mouse (SI unknown, dashed lines) in the PVN and SON. Although for different reasons, four dysregulations were shared across all four models (SON: *Fos*, *Foxp1* and *Homer1a*; PVN: *Homer1a*) and two dysregulations in at least three models (SON and PVN: *Egr1*). *Arc* was unchanged in all models in these hypothalamic structures. Data, expressed in -Δ*Ct* values, are presented as individual data, mean ± sd (**Table S5**). ANOVA followed by post-hoc tests based on estimated marginal means were performed with stars as line effect and hash as basal *vs* SI unknown 45 minutes effect (p= adjusted p-value with Benjamini-Hochberg correction). * or #p<0.05, ** or ##p<0.01, *** or ###p<0.001, **** or ####p<0.001. PVN, paraventricular nucleus; SON, supraoptic nucleus.

**Figure S14 Integration of OT and IEG mRNA levels with reciprocal social interaction in the four mouse models**

DIABLO integrated successfully qPCR values and social interaction parameters. Blue and red indicate low and high correlation, respectively. Component 1 (**A**) plots, representing correlations greater than 0.5, show a positive association of *Foxp1* in the SON and negative associations *Fos* and *Homer1a* in the CPU and PFC and *Foxp1 in the* PFC with the time spent and mean duration of nose contacts (left panel). Rearing was positively associated with *Fos* in the CPU and PFC and *Homer1a* in the PFC. *Foxp1* in the SON and nose contacts were associated with isolated mice while rearing and *Fos* and *Homer1a* in the CPU and PFC and *Foxp1* in the PFC were associated with *Oprm1* KO mice (right panel). Component 2 (**B**) shows a positive association of grooming and *Homer1a* in the PVN and a negative correlation with *Oxtr* in the PFC (left panel), which were associated with *Shank3* KO mice (right panel). In the reciprocal social interaction test (**C**), the time and the mean duration in nose contacts were increased while rearing events and cumulative time in grooming were reduced in isolated mice compared to WT animals. The time in nose contacts was reduced in *Fmr1* KO mice.

**Figure S15 *Arc* KO mice exhibit core features of ASD**

In the reciprocal social interaction test, the cumulative time in nose contacts with a genotype-, sex- and age-matched unknown mouse (**A**) was not different in *Arc* KO mice (black), compared to WT animals (gray). In motor stereotypies, the time spent in self-grooming (**B**) was increased in this model, compared to WT mice. In the three-chambered test, the cumulative time in nose contacts of *Arc* KO mice was reduced in the sociability phase (**C**), but not in the social novelty phase (**D**). Data are presented as individual data, mean ± sd. Kruskal-Wallis tests followed by Dunn post hoc tests were performed with stars as line effect and hash as chamber effect (p= adjusted p-value with Benjamini-Hochberg correction). * or # $p<0.05$, ** or ## $p<0.01$.

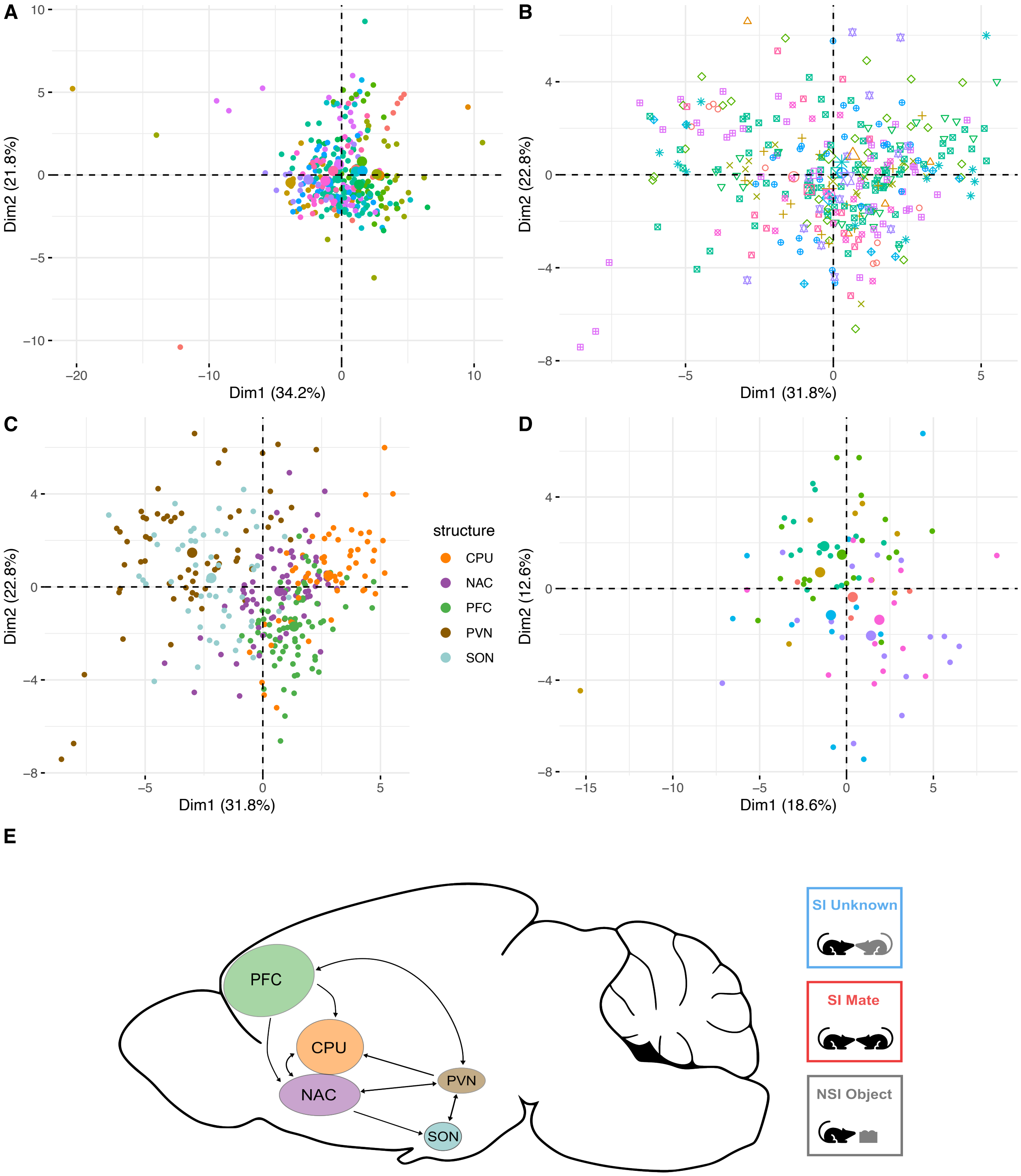

**Figure S1. Only qPCR data required batch correction over independent cohorts.** Principal Component Analysis (PCA) showed a cohort effect when analyzing the whole WT dataset (**A**). Data were corrected for batch effect and presented after correction (**B**). PCA revealed that the five brain structures formed two clusters (**C**), one with the PFC, NAC and CPU and one with the PVN and SON. PCA performed on independent cohorts of WT mice revealed no batch effect in behavior parameters (**D**). Schematic representation of the study design (**E**) with the PVN, SON, PFC, NAC and CPU in WT mice exposed to social interactions with an unknown mouse (SI unknown, blue), a cage mate (SI mate, red) or a non-social interaction with an object (NSI object, gray). CPU, caudate putamen; NAC, nucleus accumbens; PFC, prefrontal cortex; PVN, paraventricular nucleus; SON, supraoptic nucleus.

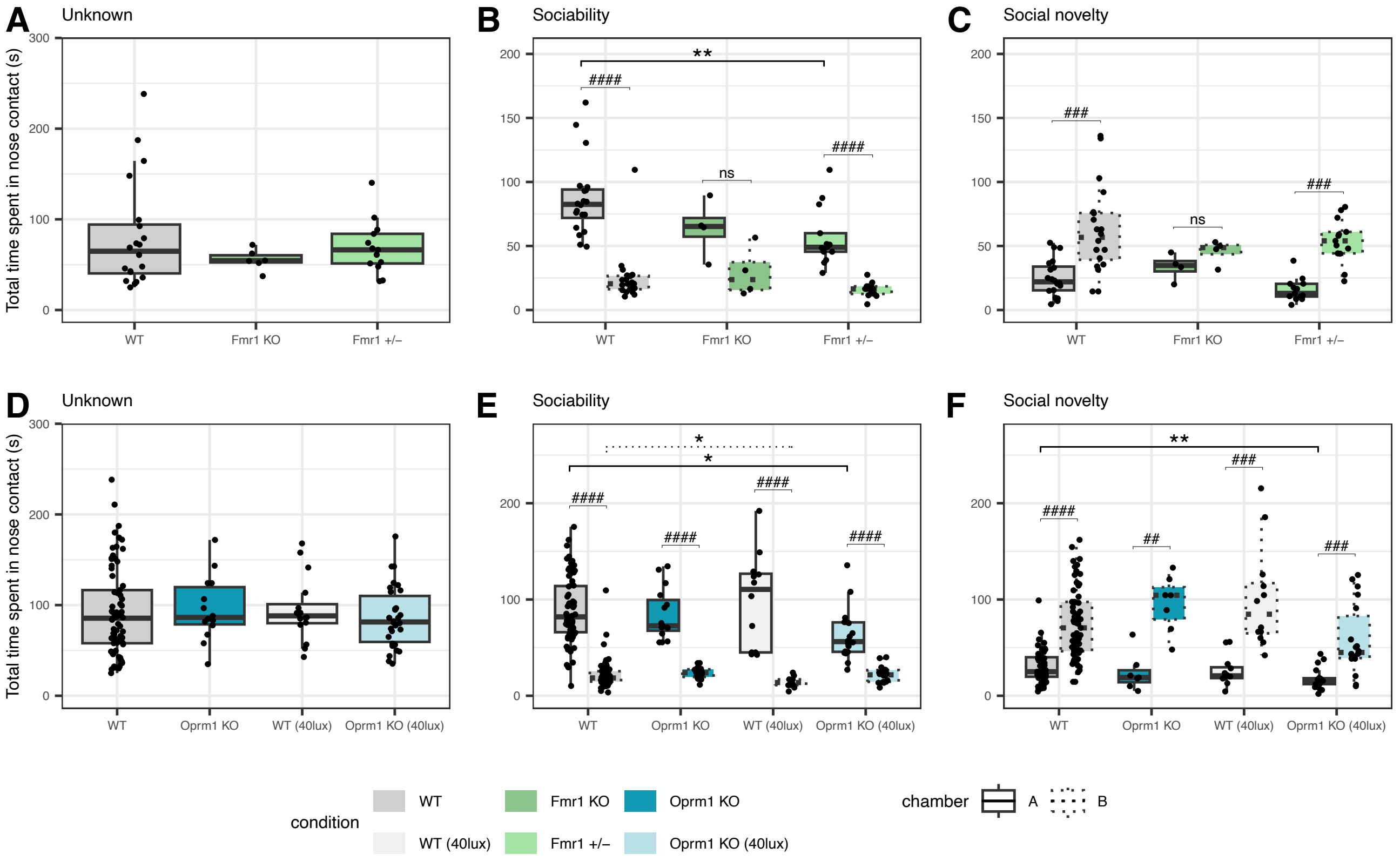

**Figure S2.** ***Fmr1*** **heterozygous females and** ***Oprm1*** **KO exposed to 40 lux display social interaction impairments.** In the reciprocal social interaction and the three-chambered tests, the cumulative time in nose contacts with a genotype-, sex- and age-matched unknown mouse (**A**, **D**) and in the sociability (**B**, **E**), the social novelty (**C**, **F**) phases of *Fmr1* KO (dark green) and heterozygous females (light green; **A**-**C**) and *Oprm1* KO (blue) and WT mice (gray), exposed to 40 (light color) or 15 lux (dark color) light intensity (**D**-**F**) are compared to WT (gray) at 15 lux. Nose contacts with an unknown conspecific are not different in any of the conditions (**A**, **D**). In the sociability phase of the three-chambered test, the time in nose contacts is reduced in *Fmr1* heterozygous females (**B**) and *Oprm1* KO (**E**) exposed to 40 lux, as well as the time in nose contact with the object in WT exposed to 40 lux. In the social novelty phase, *Fmr1* KO mice (**C**) display no preference for the novel mice (unknown WT mouse, chamber B) over the familiar mouse (already explored for 10 minutes in the precedent phase, chamber A) while *Oprm1* KO mice exposed to 40 lux (**E**) exhibit reduced time in nose contacts with the object. Data are presented as individual data, mean $\pm$ sd (**Table S3**). Kruskal-Wallis tests followed by Dunn post hoc tests were performed (p= adjusted p-value in all tests) with stars as line effect and hash as chamber effect. * or #p $<$ 0.05, ** or ##p $<$ 0.01, ###p $<$ 0.001, ####p $<$ 0.0001.

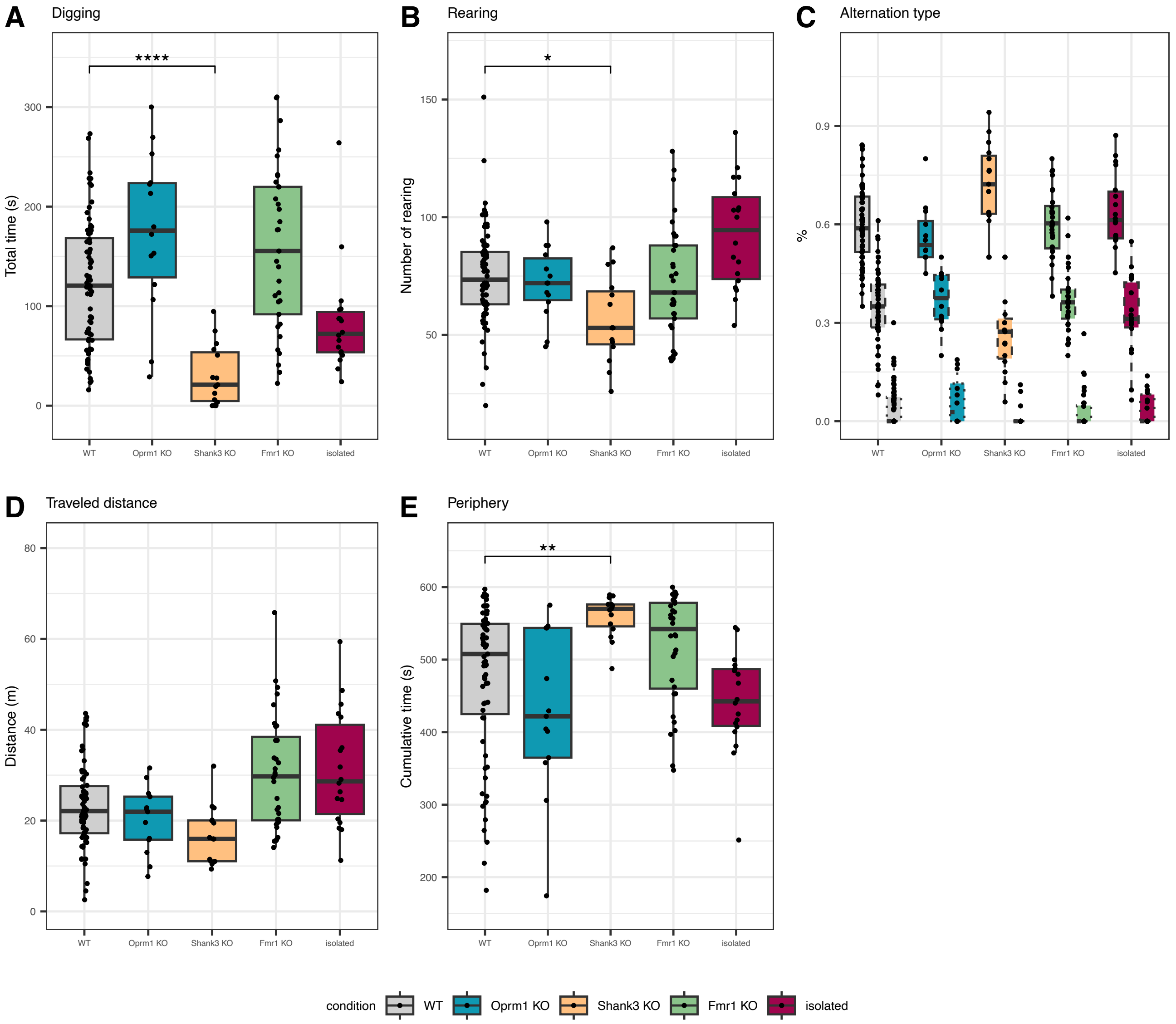

**Figure S3.** ***Shank3*** **KO mice display stereotyped and anxious-like behaviors.** In the motor stereotypy test, the time spent digging (**A**) and the number of rearing events (**B**) are reduced in *Shank3* KO mice, compared to WT. In the Y-maze test, none of the mouse lines shows impaired cognitive flexibility, as evidenced by a difference in the alternation pattern (**C**), or in the locomotion, as measured by the distance traveled in the open field (**D**). *Shank3* KO mice exhibit anxious-like behaviors as they spend more time in the periphery of the arena (**E**) than WT or other mouse models. Data are presented as individual data, mean ± sd (**Table S3**). Kruskal-Wallis tests followed by Dunn post hoc tests were performed (p= adjusted p-value in all tests). * $p < 0.05$, ** $p < 0.01$, *** $p < 0.001$, **** $p < 0.0001$.

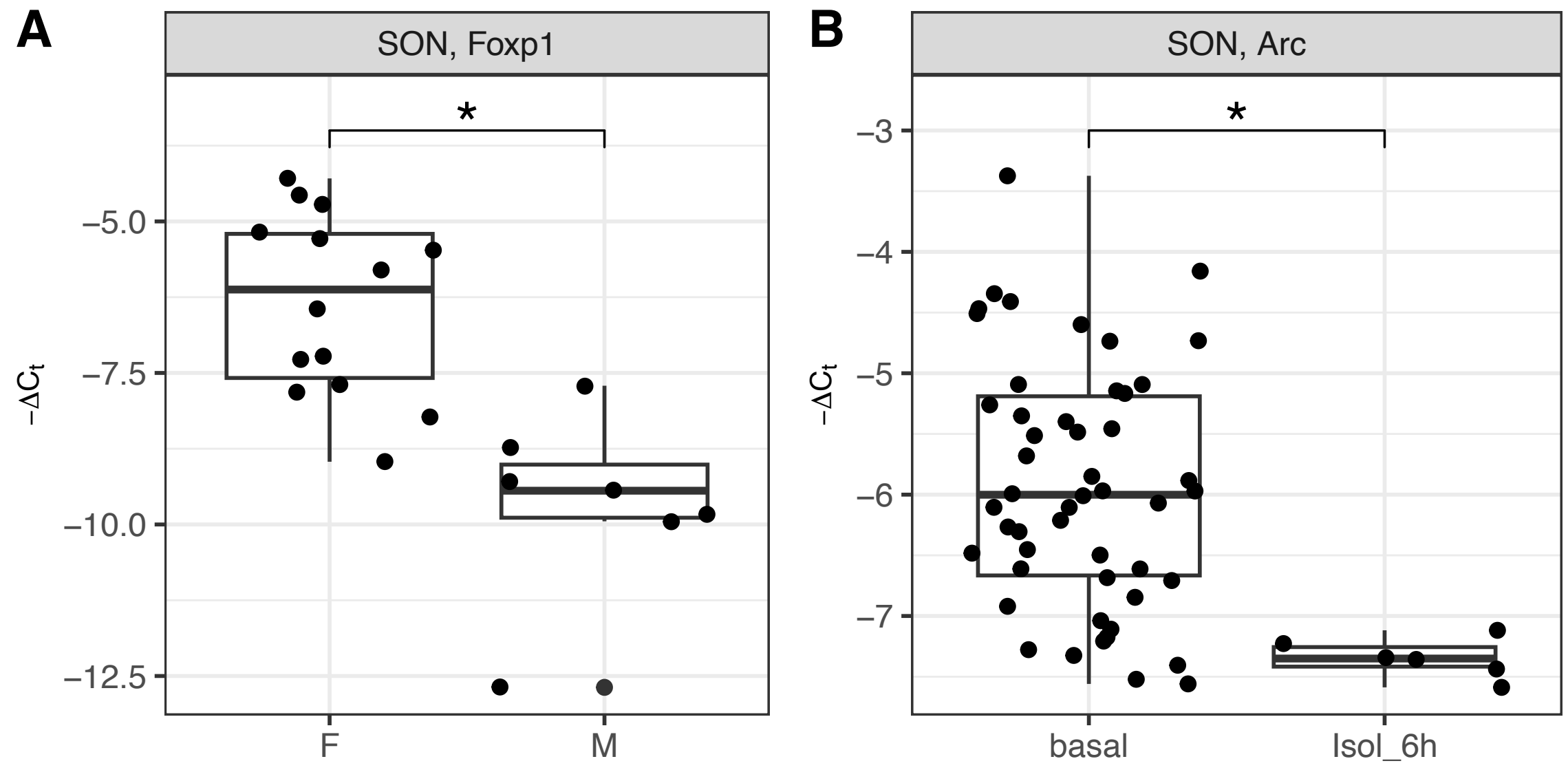

**Figure S4. Only *Foxp1* and *Arc* mRNAs in the SON of WT animals exhibit a differential pattern in males and females or exposed to 6 hours acute social isolation.** The effect of sex (**A**) and exposure to 6 hours acute isolation in the home cage (**B**) was only significant for *Foxp1* and *Arc* in the SON. *Foxp1* levels were reduced in males compared to females while *Arc* levels were decreased following acute social isolation. Data, expressed as -ΔCt values are presented as individual data, mean ± sd and detailed in Table S4. Kruskal-Wallis tests followed by Benjamini-Hochberg correction were performed with * adjusted p-value $< 0.05$. SON, supraoptic nucleus.

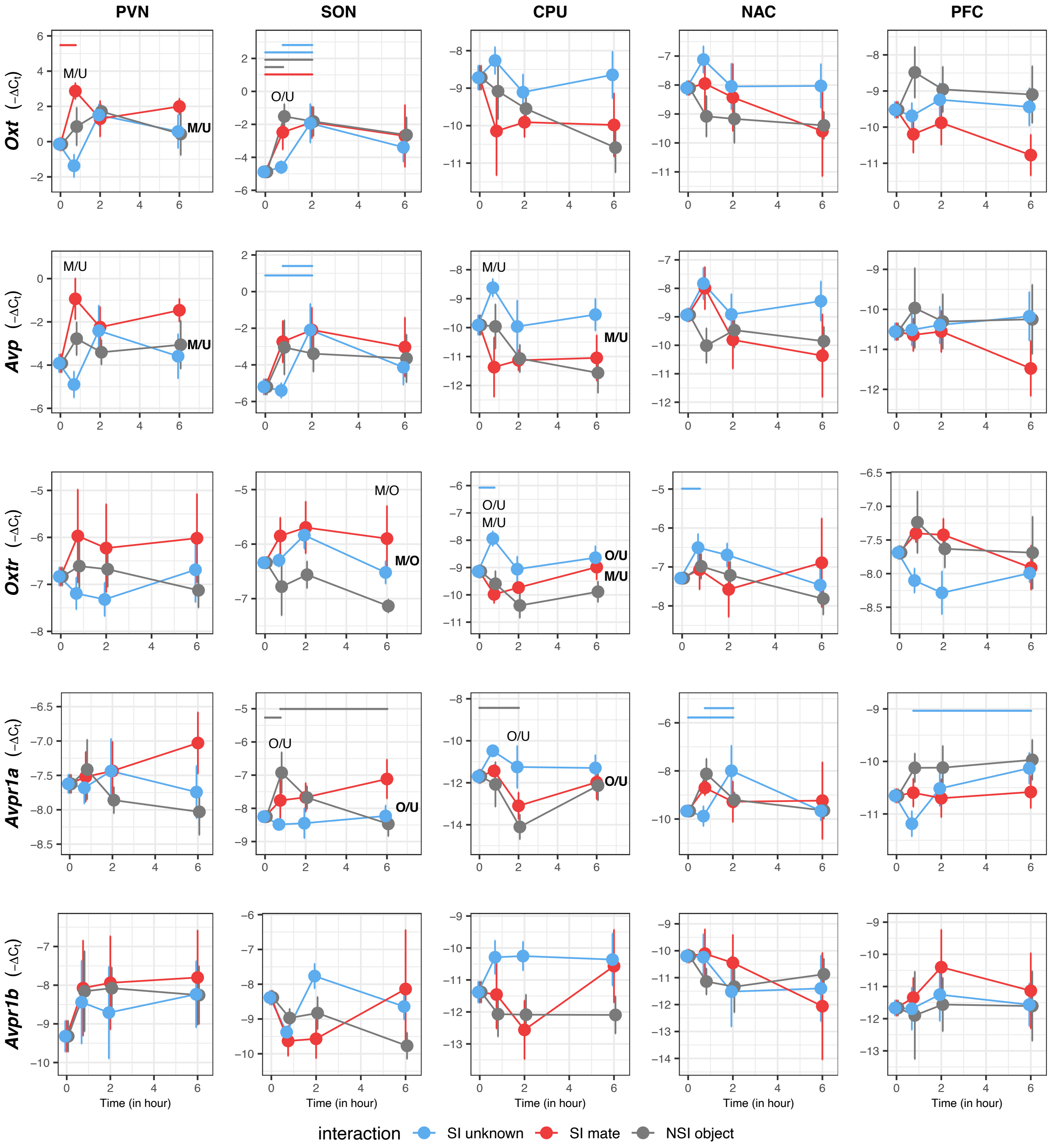

**Figure S5. Dynamic transcriptional profile of oxytocin, vasopressin and their receptor mRNAs in response to interactions.** Transcriptional profiles of *Oxt*, *Avp*, *Oxtr*, *Avpr1a* and *Avpr1b* mRNAs in WT mice are dynamic in response to social interactions with an unknown mouse (SI unknown or U; blue) or a cage mate (SI mate or M; red) or non-social interaction with an object (NSI object or O; gray) across the five structures. *Oxt* and *Avp* responded rapidly with an opposite pattern between SI unknown and SI mate in the CPU and PVN. *Oxtr* and *Avpr1a* mRNAs are dynamic in response to SI unknown, particularly in the CPU, and to a lower extent to SI mate or NSI object. *Avpr1b* mRNAs did not show any significant difference. Lines highlight a significant difference between two time points for each interaction. Letters indicate a significant difference between two interactions at a specific time point (on top of the curves) or overall pattern over time (in bold, right side of the curves). ANOVA were used to select the significant variables followed by post-hoc tests based on estimated marginal means (p= adjusted p-value in all tests). All data, expressed as -ΔCt values, mean and statistical values are detailed in **Table S4**. M, mate; O, object; U, unknown; CPU, caudate putamen; NAC, nucleus accumbens; PFC, prefrontal cortex; PVN, paraventricular nucleus; SON, supraoptic nucleus.

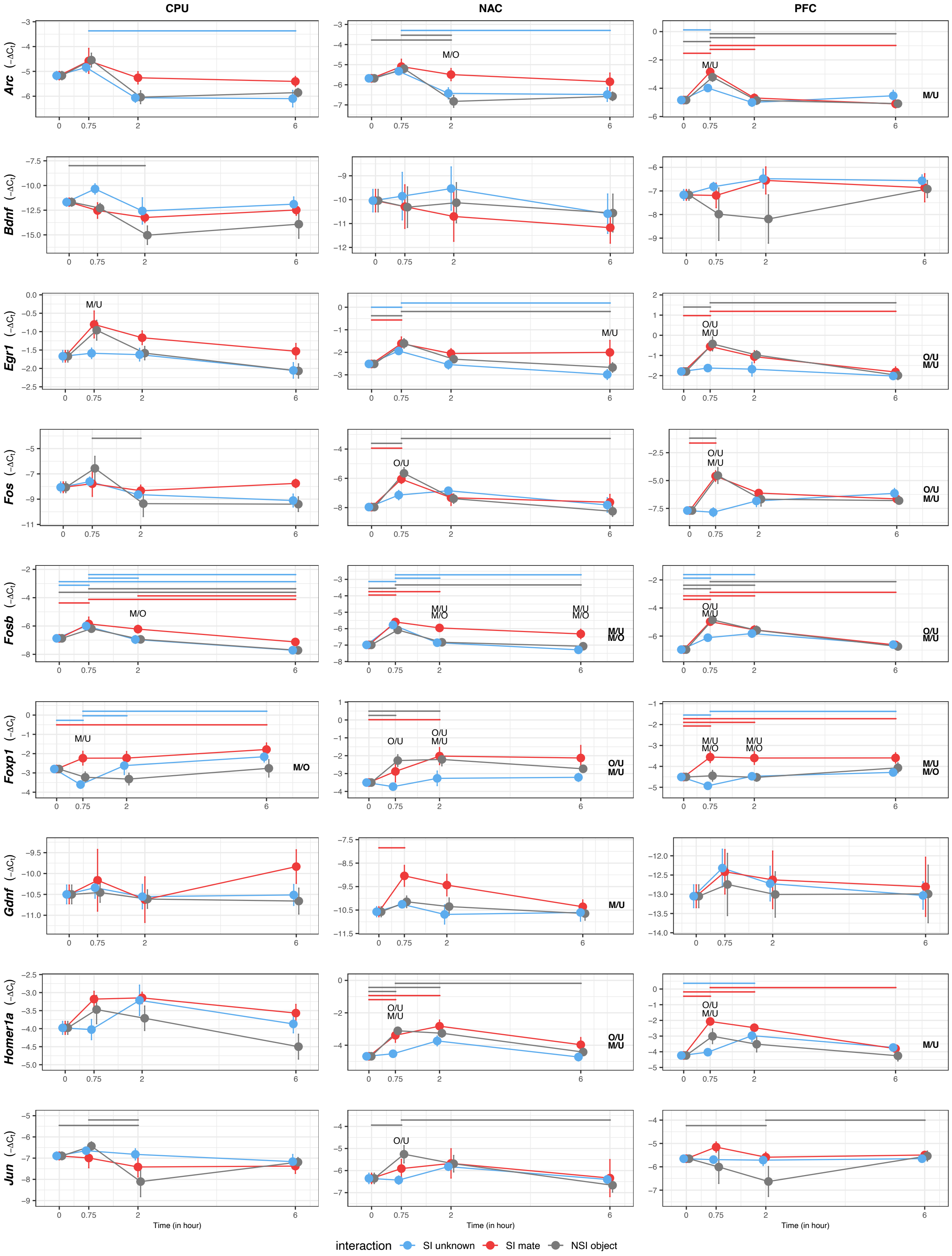

**Figure S6. Dynamic transcriptional profile of IEGs in the CPU, NAC and PFC** Transcriptional profiles of *Arc*, *Bdnf*, *Egr1*, *Fos*, *Fosb*, *Foxp1*, *Gdnf*, *Homer1a* and *Jun* show different kinetics across the CPU, NAC and PFC of WT mice at different time points (0, 0.75 or 45 minutes, 2 or 6 hours) in response to social interactions with an unknown mouse (SI unknown or U; blue), a cage mate (SI mate or M; red) or a non-social interaction with an object (NSI object or O; gray). *Arc*, *Egr1*, *Fos*, *Foxp1* and *Homer1a* showed the most significant differences between social and non-social interactions. Lines highlight a significant difference between two time points for each interaction. Letters indicate a significant difference between two interactions at a specific time point (on top of the curves) or overall pattern over time (in bold, right side of the curves). ANOVA were used to select the significant variables followed by post-hoc tests based on estimated marginal means (p= adjusted p-value in all tests). All data, expressed as -ΔCt values, mean and statistical values are detailed in the **Table S4**. PFC, prefrontal cortex; NAC, nucleus accumbens; CPU, caudate putamen.

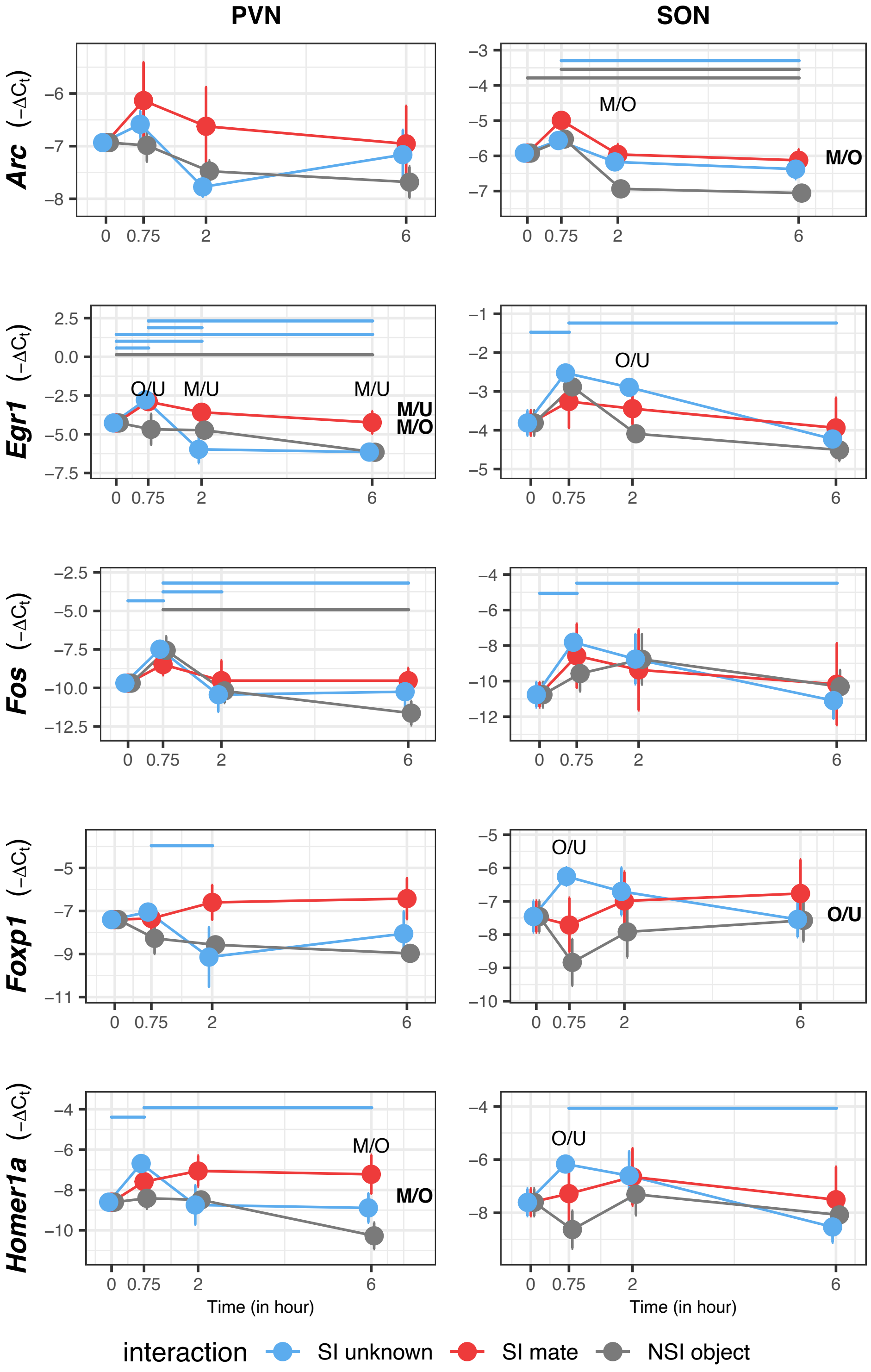

**Figure S7. Dynamic transcriptional profile of IEGs in the PVN and SON** Transcriptional profiles of *Arc*, *Egr1*, *Fos*, *Fosb*, *Foxp1* and *Homer1a* show different kinetics across the PVN and SON of WT mice at different time points (0, 0.75 or 45 minutes, 2 or 6 hours) in response to social interactions with an unknown mouse (SI unknown or U; blue), a cage mate (SI mate or M; red) or a non-social interaction with an object (NSI object or O; gray). Except *Arc* for SI mate, *Egr1*, *Fos*, *Foxp1* and *Homer1a* exhibit the most significant differences for SI unknown in these hypothalamic structures. Lines highlight a significant difference between two time points for each interaction. Letters indicate a significant difference between two interactions at a specific time point (on top of the curves) or overall pattern over time (in bold, right side of the curves). ANOVA were used to select the significant variables followed by post-hoc tests based on estimated marginal means (p= adjusted p-value in all tests). All data, expressed as -ΔCt values, mean and statistical values are detailed in the **Table S4**. PVN, paraventricular nucleus; SON, supraoptic nucleus.

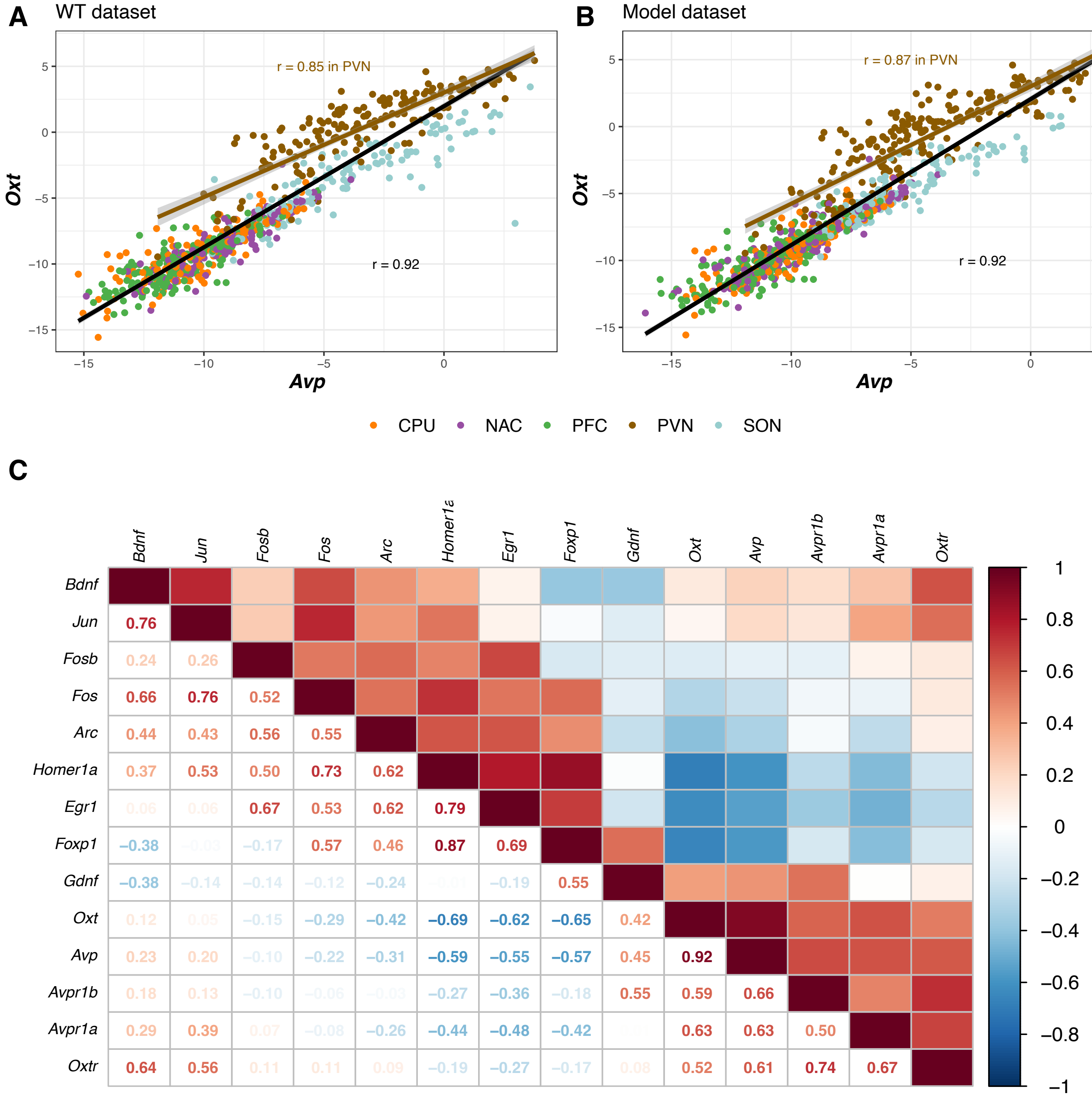

**Figure S8.** ***Oxt* and *Avp* mRNAs are highly correlated.** *Oxt* and *Avp* mRNA levels are highly correlated in WT (**A**) and the four mouse models (**B**), reaching 92% across the NAC, CPU, SON and PFC and 85 and 87%, respectively, in the PVN. Correlation plot between the different mRNAs in WT animals among the OT and IEG families (**C**) reveals in blue and red, low and high levels of correlation, respectively. In addition to *Oxt* and *Avp*, *Homer1a* is the most positively correlated to *Foxp1* and *Egr1* mRNAs, followed by *Fos* with *Jun* and *Homer1a*, *Bdnf* and *Jun*, and *Oxtr* with *Avpr1b*. Interestingly, *Homer1a*, *Foxp1* and *Egr1* were negatively correlated to *Oxt* and *Avp*. PFC, prefrontal cortex; NAC, nucleus accumbens; CPU, caudate putamen; PVN, paraventricular nucleus; SON, supraoptic nucleus.

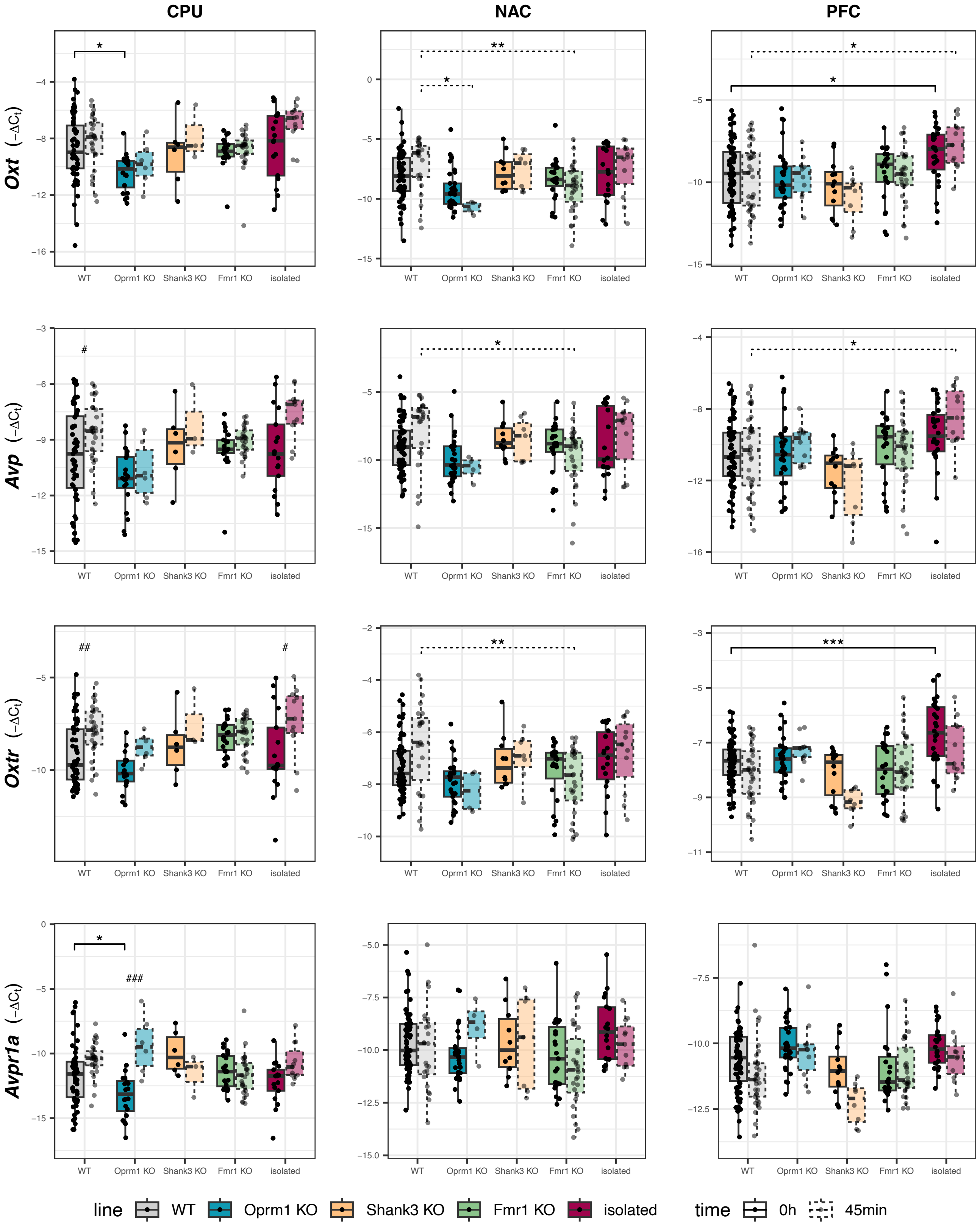

**Figure S9. OT family dysregulations in the CPU, NAC and PFC of mouse models.** Dysregulations of *Avp*, *Avpr1a*, *Oxt* and *Oxtr* mRNAs within the OT family in *Oprm1* KO (blue), *Shank3* KO (yellow), *Fmr1* KO (green) and isolated (burgundy) mice under basal conditions (solid lines) or 45 minutes following social interaction with an unknown mouse (SI unknown, dashed lined) in the CPU, NAC and PFC were compared to levels in WT mice. *Avp* mRNA levels were not induced in response to SI unknown across the four mouse models, as well as *Oxtr* levels in *Oprm1*, *Shank3* and *Fmr1* KO mice, compared to WT animals. *Oxt* levels after SI unknown are lower in the NAC of *Fmr1* and *Oprm1* KO mice. *Avp*, *Oxt* and *Oxtr* mRNAs were up-regulated in the PFC of isolated mice. Data, expressed as -ΔCt values, are presented as individual data, mean ± sd and detailed in **Table S5**. ANOVA followed by post-hoc tests based on estimated marginal means were performed with stars as line effect and hash as basal vs SI unknown 45 minutes effect (p= adjusted p-value with Benjamini-Hochberg correction). * or #p $< 0.05$, ** or ##p $< 0.01$, *** or ###p $< 0.001$. PFC, prefrontal cortex; NAC, nucleus accumbens; CPU, caudate putamen.

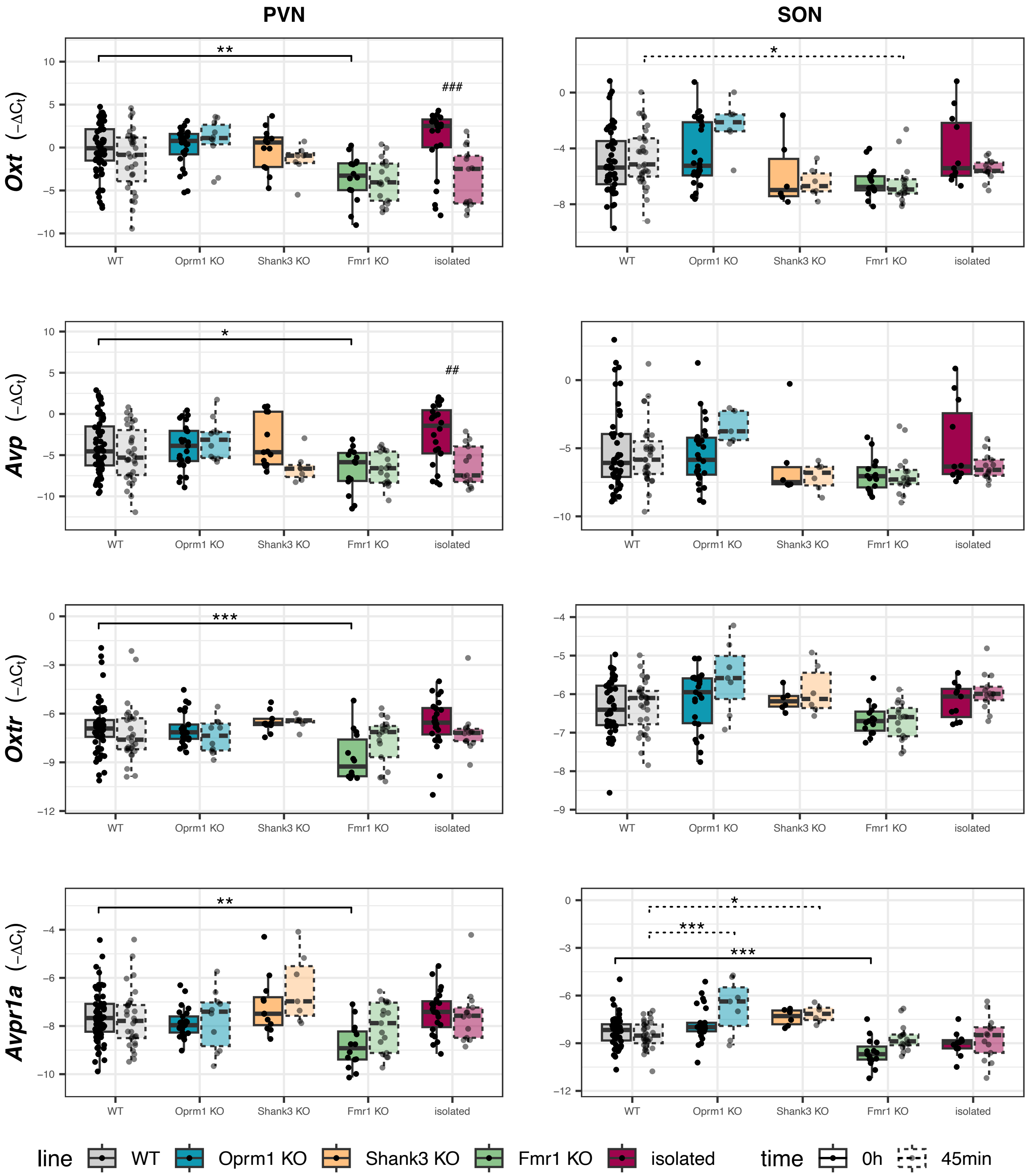

**Figure S10. OT family dysregulations in the PVN and SON of mouse models.** Dysregulations of *Oxt*, *Avp*, *Oxtr*, *Avpr1a* mRNAs within the OT family in *Oprm1* KO (blue), *Shank3* KO (yellow), *Fmr1* KO (green) and isolated (burgundy) mice under basal conditions (solid lines) or 45 minutes following social interaction with an unknown mouse (SI unknown, dashed lines) in the PVN and SON were compared to WT mice. All four mRNAs are down-regulated in the PVN of *Fmr1* KO mice, as well as *Oxt* and *Avpr1a* in the SON, compared to WT animals. *Avpr1a* levels after SI unknown are lower in the SON of *Shank3* and *Oprm1* KO mice. Data, expressed as -ΔCt values, are presented as individual data, mean ± sd and detailed in **Table S5**. ANOVA followed by post-hoc tests based on estimated marginal means were performed with stars as line effect and hash as basal vs SI unknown 45 minutes effect (p= adjusted p-value with Benjamini-Hochberg correction). * or #p $<$ 0.05, ** or ##p $<$ 0.01, *** or ###p $<$ 0.001. PVN, paraventricular nucleus; SON supraoptic nucleus.

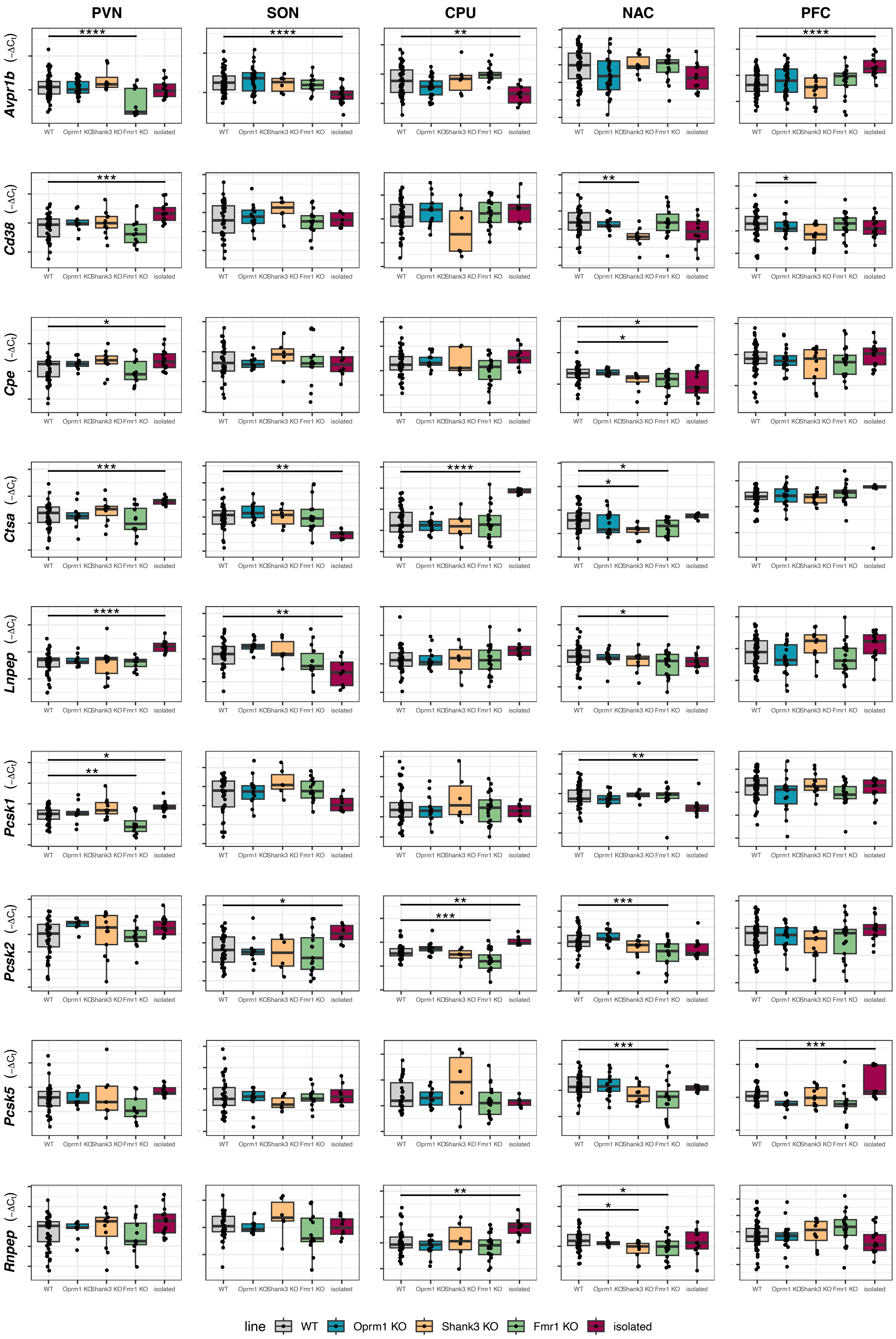

**Figure S11. mRNA levels in the extended OT family under basal conditions in mouse models.** Dysregulations of receptor (*Avpr1b*) and mRNAs involved in OT secretion (*Cd38*), peptide biosynthesis (*Pcsk1*, *Pcsk2*, *Pcsk5* and *Cpe*) and degradation (*Lnpep*, *Ctsa* and *Rnpep*) in *Oprm1* KO (blue), *Shank3* KO (yellow), *Fmr1* KO (green) and isolated (burgundy) mice compared to WT mice are presented under basal conditions. Data are presented as individual data, mean ± sd (**Table S5**). Linear, followed by post-hoc tests. * p adjusted $< 0.05$, ** $< 0.01$, *** $< 0.001$, **** $< 0.0001$.

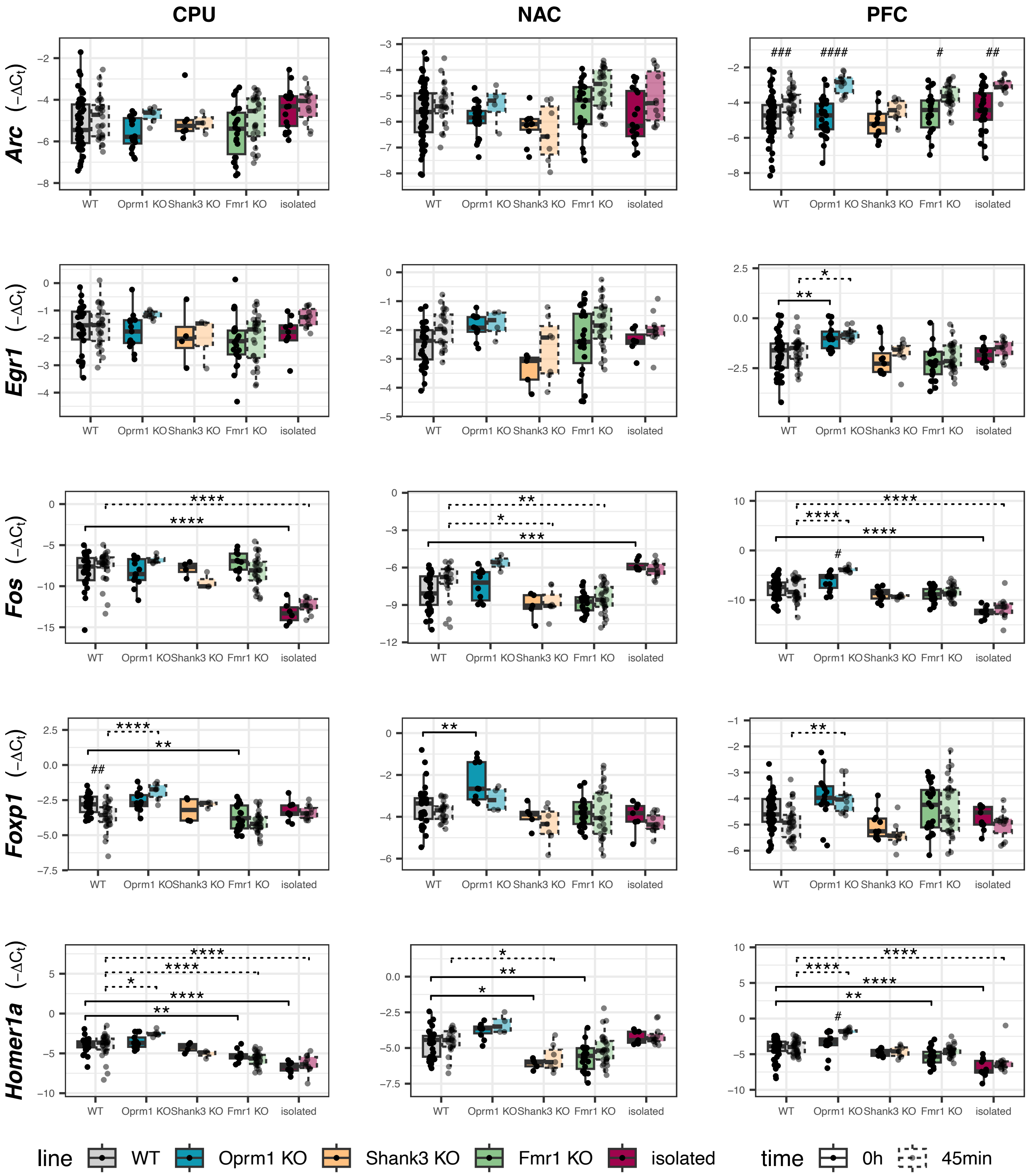

**Figure S12. IEG dysregulations in the CPU, NAC and PFC of mouse models.** Dysregulations of *Arc*, *Egr1*, *Fos*, *Foxp1* and *Homer1a* among the IEG family in *Oprm1* KO (blue), *Shank3* KO (yellow), *Fmr1* KO (green) and isolated (burgundy) mice were compared to WT mice under basal conditions (solid lines) or 45 minutes following social interaction with an unknown mouse (SI unknown, dashed lines) in the CPU, NAC and PFC. Although for distinct reasons, one dysregulation was shared across all four models (CPU: *Foxp1*) and four dysregulations were common between *Fmr1* and *Shank3* KO mice (NAC: *Fos*, *Homer1a*) or isolated mice (CPU and PFC: *Homer1a*). Nine dysregulations were specific to a model. Only in the PFC, *Arc* was not significantly induced in *Shank3* KO mice, compared to WT mice. Data, expressed in -ΔCt values, are presented as individual data, mean ± sd (**Table S5**). ANOVA followed by post-hoc tests based on estimated marginal means were performed with stars as line effect and hash as basal vs SI unknown 45 minutes effect (p= adjusted p-value with Benjamini-Hochberg correction). * or #p $<$ 0.05, ** or ##p $<$ 0.01, *** or ###p $<$ 0.001, **** or ####p $<$ 0.001. CPU, caudate putamen; NAC, nucleus accumbens; PFC, prefrontal cortex.

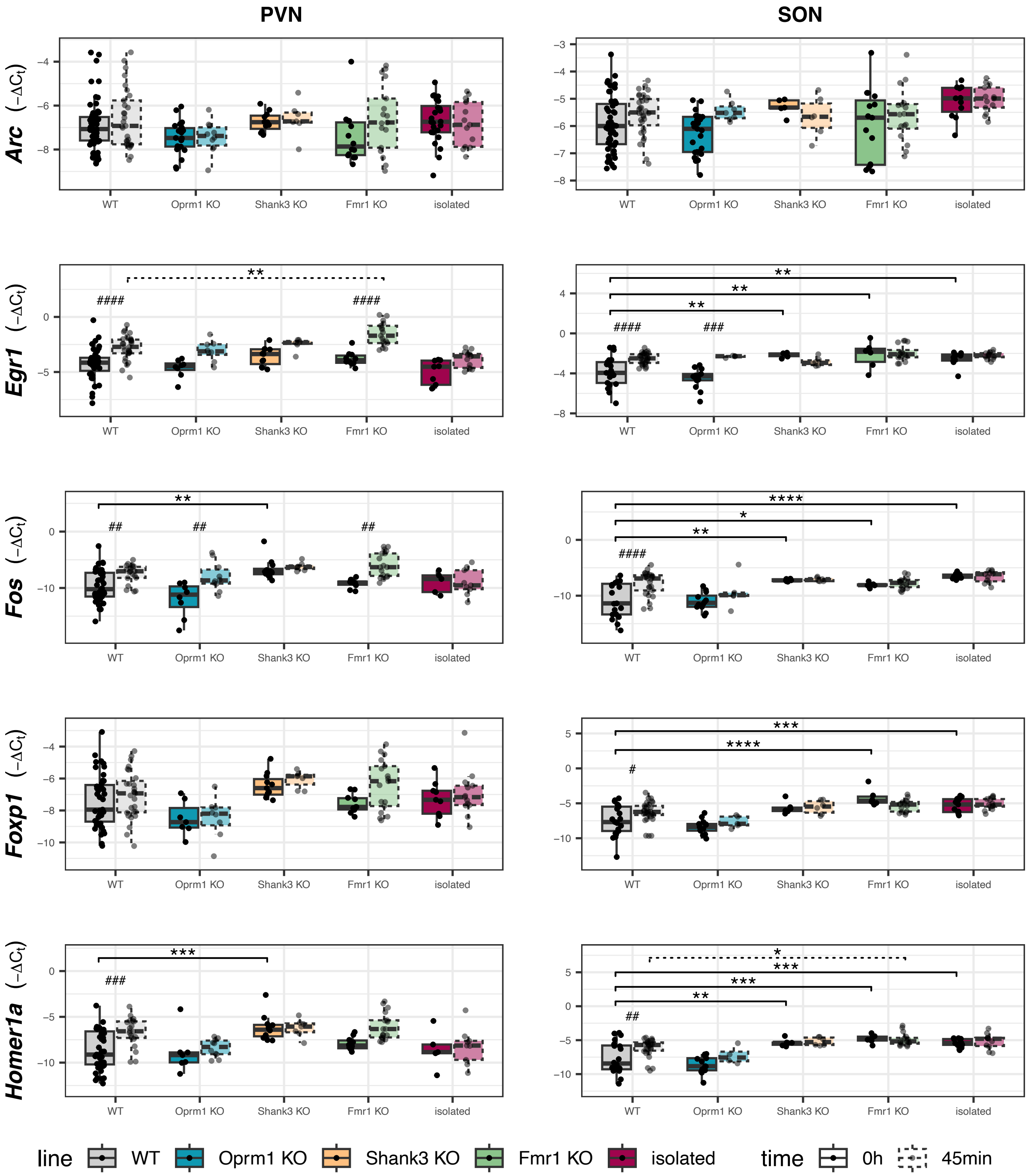

**Figure S13. IEG dysregulations in the PVN and SON of mouse models.** Dysregulations of *Arc*, *Egr1*, *Fos*, *Foxp1* and *Homer1a* among the IEG family in *Oprm1* KO (blue), *Shank3* KO (yellow), *Fmr1* KO (green) and isolated (burgundy) mice were compared to WT mice under basal conditions (solid lines) or 45 minutes following social interaction with an unknown mouse (SI unknown, dashed lines) in the PVN and SON. Although for different reasons, four dysregulations were shared across all four models (SON: *Fos*, *Foxp1* and *Homer1a*; PVN: *Homer1a*) and two dysregulations in at least three models (SON and PVN: *Egr1*). *Arc* was unchanged in all models in these hypothalamic structures. Data, expressed in -ΔCt values, are presented as individual data, mean ± sd (**Table S5**). ANOVA followed by post-hoc tests based on estimated marginal means were performed with stars as line effect and hash as basal vs SI unknown 45 minutes effect (p= adjusted p-value with Benjamini-Hochberg correction). * or #p $<$ 0.05, ** or ##p $<$ 0.01, *** or ###p $<$ 0.001, **** or ####p $<$ 0.001. PVN, paraventricular nucleus; SON, supraoptic nucleus.

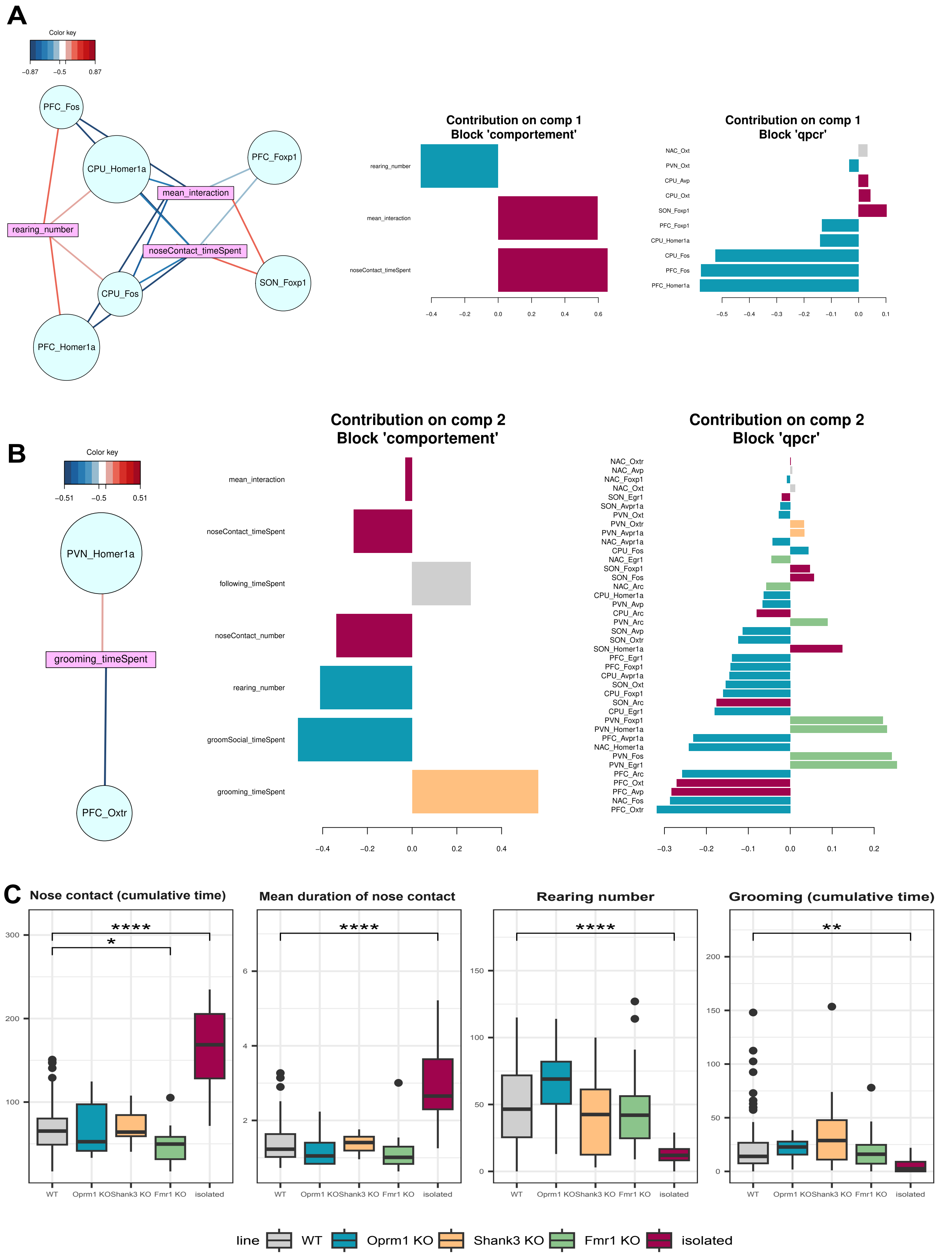

**Figure S14. Integration of OT and IEG mRNA levels with reciprocal social interaction in the four mouse models.** DIABLO integrated successfully qPCR values and social interaction parameters. Blue and red indicate low and high correlation, respectively. Component 1 (**A**) plots, representing correlations greater than 0.5, show a positive association of *Foxp1* in the SON and negative associations *Fos* and *Homer1a* in the CPU and PFC and *Foxp1* in the PFC with the time spent and mean duration of nose contacts (left panel). Rearing was positively associated with *Fos* in the CPU and PFC and *Homer1a* in the PFC. *Foxp1* in the SON and nose contacts were associated with isolated mice while rearing and *Fos* and *Homer1a* in the CPU and PFC and *Foxp1* in the PFC were associated with *Oprm1* KO mice (right panel). Component 2 (**B**) shows a positive association of grooming and *Homer1a* in the PVN and a negative correlation with *Oxtr* in the PFC (left panel), which were associated with *Shank3* KO mice (right panel). In the reciprocal social interaction test (**C**), the time and the mean duration in nose contacts were increased while rearing events and cumulative time in grooming were reduced in isolated mice compared to WT animals. The time in nose contacts was reduced in *Fmr1* KO mice.

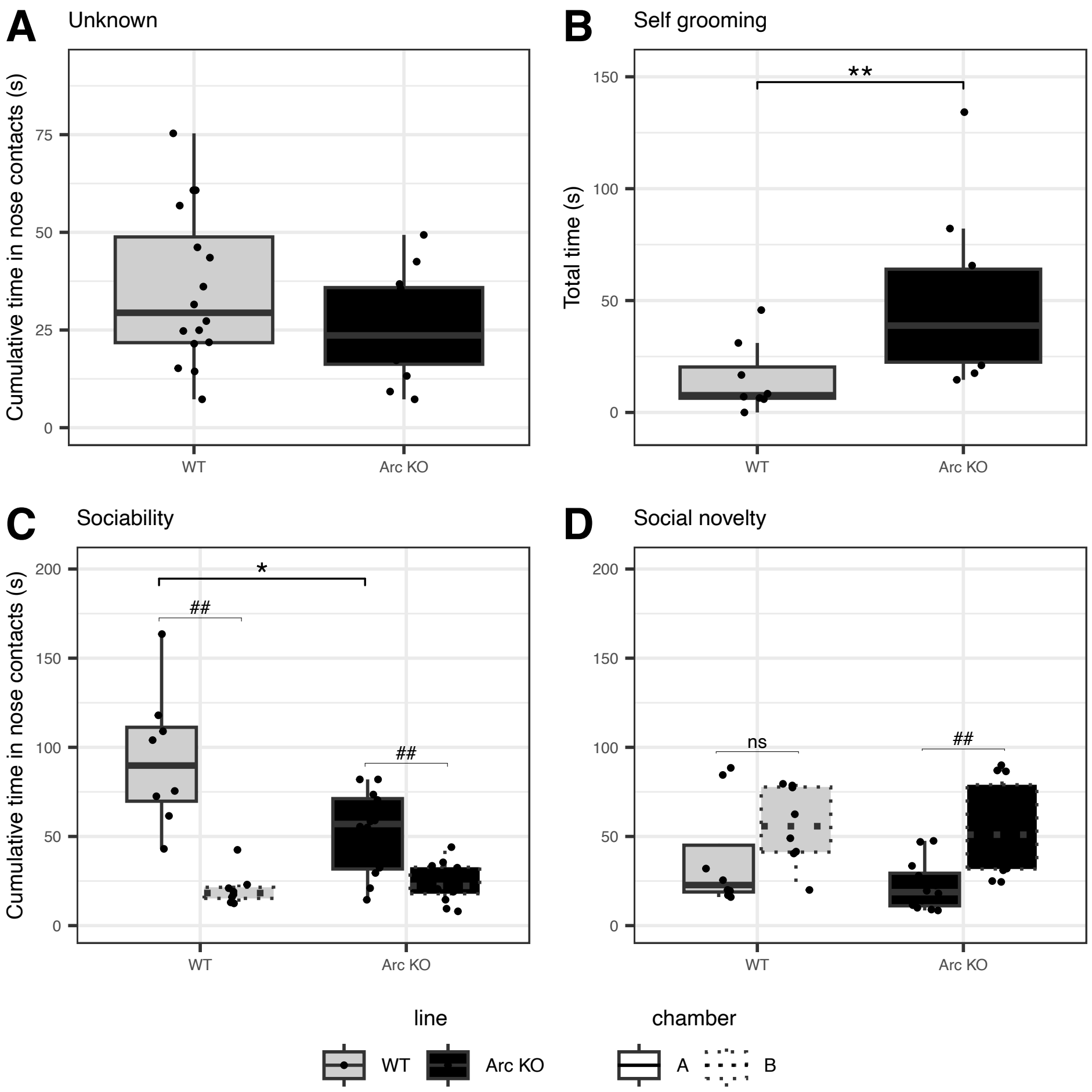

**Figure S15.** ***Arc*** **KO mice exhibit core features of ASD.** In the reciprocal social interaction test, the cumulative time in nose contacts with a genotype-, sex- and age-matched unknown mouse (**A**) was not different in *Arc* KO mice (black), compared to WT animals (gray). In motor stereotypies, the time spent in self-grooming (**B**) was increased in this model, compared to WT mice. In the three-chambered test, the cumulative time in nose contacts of *Arc* KO mice was reduced in the sociability phase (**C**), but not in the social novelty phase (**D**). Data are presented as individual data, mean ± sd. Kruskal-Wallis tests followed by Dunn post hoc tests were performed with stars as line effect and hash as chamber effect (p= adjusted p-value with Benjamini-Hochberg correction). * or # $p < 0.05$, ** or ## $p < 0.01$.